%% file: fourier_v2.tex
\documentclass[a4paper, 10pt] {article}
\usepackage{jheppub}
\include{preamblejhep}

\usepackage{comment}

\title{Quantization maps, algebra representation and non-commutative Fourier transform for Lie groups\footnote{The following article appeared in J.\@ Math.\@ Phys.\@ \textbf{54}, 083508 (2013) and may be found at \href{http://dx.doi.org/10.1063/1.4818638}{doi:10.1063/1.4818638}. Copyright 2013 \href{http://www.aip.org/}{American Institute
of Physics}. This article may be downloaded for personal use only. Any other use requires prior permission of the author and the American Institute of Physics.}}
\author[a]{Carlos Guedes,}
\author[a]{Daniele Oriti,}
\author[a,b]{Matti Raasakka}
\affiliation[a]{Max Planck Institute for Gravitational Physics (Albert Einstein Institute),\\Am M\"uhlenberg 1, 14476 Golm, Germany}
\affiliation[b]{LIPN, Institut Galil\'ee, Universit\'e Paris-Nord,\\ 99 av. Clement, 93430 Villetaneuse, France\\}
\emailAdd{carlos.guedes@aei.mpg.de} 
\emailAdd{daniele.oriti@aei.mpg.de} 
\emailAdd{matti.raasakka@lipn.univ-paris13.fr}

\abstract{The phase space given by the cotangent bundle of a Lie group appears in the context of several models for physical systems. A representation for the quantum system in terms of non-commutative functions on the (dual) Lie algebra, and a generalized notion of (non-commutative) Fourier transform, different from standard harmonic analysis, has been recently developed, and found several applications, especially in the quantum gravity literature. We show that this algebra representation can be defined on the sole basis of a quantization map of the classical Poisson algebra, and identify the conditions for its existence. In particular, the corresponding non-commutative star-product carried by this representation is obtained directly from the quantization map via deformation quantization. We then clarify under which conditions a unitary intertwiner between such algebra representation and the usual group representation can be constructed giving rise to the non-commutative plane waves and, consequently, the non-commutative Fourier transform. The compact groups $\U(1)$ and $\SU(2)$ are considered for different choices of quantization maps, such as the symmetric and the Duflo map, and we exhibit the corresponding star-products, algebra representations and non-commutative plane waves.}

\begin{document}
\maketitle
\section{Introduction}
In ordinary quantum mechanics of a point particle on flat space, we can either choose to represent our wave functions in the position representation, that is, realizing the Hilbert space of the system as  $L^2$ functions on the configuration space, or in the momentum representation, given again by $L^2$ functions on the cotangent space. These two realizations can be independently defined, once a quantization map of the classical Poisson algebra of observables has been chosen. On a Euclidean space the usual Fourier transform gives a map between both representations, i.e., between the two $L^2$ spaces, relating them self-dually. Explicitly, for $\psi\in L^2(\Rl^d)$, the Fourier transform is given by
\ba
\tilde{\psi}(\vec{p})=\int_{\Rl^d}{\d^d x}\, e^{-i\vec{p}\cdot\vec{x}}\,\psi(\vec{x})\;\in L^2(\Rl^d)\,, \nonumber
\ea
where $e^{-i\vec{p}\cdot\vec{x}}$ are unitary irreducible representations of the group of translations in $\Rl^d$, and $\vec{x},\vec{p}$ vectors in $\Rl^d$. Thus, in the flat case, points on the cotangent (momentum) space are in one-to-one correspondence with unitary irreducible representations of the translational symmetry group of the configuration space.

For a generic curved manifold, a momentum representation in terms of $L^2$ functions on its cotangent space cannot be defined, in absence of symmetries, nor a notion of Fourier transform. On the other hand, for symmetric spaces and, in particular, for Lie groups the notion of Fourier transform can be generalized as an expansion in terms of unitary irreducible representations of the same group, acting transitively on the configuration manifold. More precisely, for any locally compact group $G$, the Fourier transform is defined as the unitary map between $L^2(G)$ and $L^2(\widehat{G})$, where $\widehat{G}$, the Pontryagin dual of $G$, denotes the set of equivalence classes of unitary irreducible representations of $G$. Harmonic analysis is, indeed, a very useful tool in quantum mechanics, quantum field theory in curved spaces, and quantum gravity.

However, some of the nice features of the usual momentum representation and of the usual Fourier transform are inevitably lost. When considering a physical system whose configuration space is a Lie group $G$ (e.g., a particle on a 3-sphere described by $\SU(2)$), the momentum space coincides with the dual of the Lie algebra $\fg^*$, which in general differs from $\widehat{G}$. For example, for $\SU(2)$, $\widehat{\SU(2)}=\Nl_0/2$, while $\su(2)^*\simeq \Rl^3$. That is, the Pontryagin dual is a very different object from the cotangent space of a configuration space, coinciding only in very special cases, as $G=\Rl^d$ above.
Therefore, the dual representation obtained from harmonic analysis is not in terms of (generalized) functions of momenta, i.e., functions on the Lie algebra. This implies that one is bound to lose contact with the classical theory, at least at the formal level, when working with quantum observables that are functions of the momenta. Of course, the same physical information can be recovered in any representation of the quantum system, but one would like to maintain a closer formal resemblance with the classical quantities, to help maintaining also a closer contact with the underlying physics. In particular, several quantum gravity approaches, most notably loop quantum gravity \cite{TTbook, CRbook, Rovelli:2011eq}, spin foam models \cite{Perez:2012wv} and group field theories \cite{Oriti:2011jm, Oriti:2006se, Baratin:2011aa}, work with an underlying classical phase space based on the cotangent bundle over a Lie group (either $\SU(2)$ or the Lorentz group $\SL(2,\Cl)$). While the group elements encode the degrees of freedom of the gravitational connection, the elements of the Lie algebra are related directly to the triad field, thus to the metric itself. A representation, which makes directly use of functions of such Lie algebra elements, would then bring the geometric aspects of the theory to the forefront.    

Such Lie algebra representation has been proposed in the quantum gravity context (where it also goes under the name of flux representation) and its development and application is now a growing area of research \cite{Freidel:2005bb, Freidel:2005ec, JoungMouradNoui, Baratin:2010wi, Baratin:2011tx, Baratin:2011hp, Baratin:2011tg, Oriti:2011ug, DanieleMatti, Dupuis:2011fx}. However, it has been used, up to now, as a derived product of the usual group representation, and obtained from a non-commutative Fourier transform whose mathematical basis has remained only partially explored, and which has still a certain flavour of arbitrariness in its defining details (e.g., plane waves and star-products).

The goals of this article are the following. First of all, we want to show that the algebra representation can be defined independently of the group representation, on the sole basis of the choice of a quantization map of the classical Poisson algebra, and identify more clearly the conditions for its existence. Second, we want to clarify under which conditions a unitary map between such an algebra representation (assuming it exists) and the usual group representation can be constructed, that is, characterize the non-commutative Fourier transform together with the corresponding non-commutative plane waves. In looking to the above, we try to work with as general a Lie group $G$ as possible. Third, we want to consider specific and interesting choices of quantization maps and Lie groups, and exhibit the corresponding star-products, algebra representations and non-commutative plane waves. On the one hand, we prove with these examples the non-emptiness of the definitions provided together with the existence of their algebra representation and of their non-commutative Fourier transforms. On the other hand, the results of specific quantization maps can find direct applications, as we discuss in the following, to quantum gravity models. In particular, we identify the non-commutative plane waves and a star-product for the Duflo map --- a special case of the Kontsevich star-product ---, which has been suggested to be useful in several quantum gravity contexts \cite{Alekseev:2000hf, Sahlmann:2011rv, Sahlmann:2011uh, Noui:2011im}. 

The construction we present in this article extends earlier work on the non-commutative Fourier transform by several authors. The concept arose originally in considerations of the phase space structure of 3d Euclidean quantum gravity models. The earliest notion (to our knowledge) of a non-commutative Fourier transform for the group $\SU(2)$ appeared in a paper by Schroers \cite{Schroers} (see also \cite{Majid:2008iz} by Schroers \& Majid), where the construction is based on the duality structure of the quantum double $\textrm{DSU}(2)$, which is introduced as a quantization of the classical phase space $\textrm{ISO}(3)$. Later, more explicit notions of what became to be called `group' Fourier transform were introduced, first for the group $\SO(3)$ by Freidel \& Livine\cite{Freidel:2005bb}, and later extended to $\SU(2)$ and related to the quantum group Fourier transform by Freidel \& Majid \cite{Freidel:2005ec}, Joung, Mourad \& Noui \cite{JoungMouradNoui} and Dupuis, Girelli \& Livine \cite{Dupuis:2011fx}, each in their own different ways. See also \cite{Rosa:2012pr, GraciaBondia:2001ct}. To a certain extent, our construction in this paper can be considered as yet another extension of the original concept of Freidel \& Livine \cite{Freidel:2005bb} to more general classes of non-commutative structures and Lie groups. However, it derives from the canonical structures of the classical phase space, the cotangent bundle of $G$, of the quantization map applied to it, and of the corresponding quantum observable algebra. Thus, it also provides a better general understanding of the relation of the non-commutative Fourier transform to these fundamental underlying structures.

For other directions to Fourier analysis on Lie groups, let us in particular point to the extensive work on the Kirillov orbit method \cite{kirillov}, subsequent (Fourier) analysis based on the decomposition of $\widehat{G}$ into orbits in $\fg^*$ \cite{wildberger}, and the Helgason Fourier transform \cite{helgason} for further reference.\\

Let us summarize our results. The starting point is the Poisson algebra associated to the cotangent bundle of a Lie group $G$, taken to be $\cp_G=(C^\infty(G\times\fg^*),\{\cdot,\cdot\}, \cdot)$ with canonical symplectic structure $\{\cdot,\cdot\}$, and pointwise multiplication $\cdot$. Canonical quantization of (a suitable subalgebra of) $\cp_G$ gives an abstract operator $^*$-algebra $\fA$ endowed with natural Hopf algebra structures. A representation of $\fA$ on the Hilbert space $L^2(G)$ of square-integrable functions on $G$ (with respect to the Haar measure $\d g$) is straightforwardly available as any set of coordinates on $G$ form (in an implicit sense given below) a simultaneously diagonizable maximal abelian subalgebra of self-adjoint operators. This provides the \emph{group representation}. A definition of a dual \emph{algebra representation} of $\fA$ in terms of a function space we denote by $L^2_\star(\fg^*)$ is made possible by introducing a star-product $\star$ in the sense of deformation quantization \cite{defquant}, depending only on the chosen quantization map from $\cp_G$ to $\fA$. In particular, the inner product in this Hilbert space is the $L^2$ inner product with respect to a star-product $\star_p$ (and the Lebesgue measure $\d^d X$ on $\fg^*$), which is the deformation quantization star-product $\star$ amended with a projection that accounts for the compact subgroups of $G$; namely, $\langle f, g\rangle = \int \frac{\d^d X}{(2\pi)^d}\ \overline{f} \star_p g$. We show under which conditions on the star-product, such algebra representation can be defined. The \emph{non-commutative Fourier transform} is then shown to arise as the intertwiner between these two representations. For $\psi\in L^2(G)$ and $\tilde{\psi}\in L^2_\star(\fg^*)$, the non-commutative Fourier transform $\cf$ and its inverse $\cf^{-1}$ are determined to be
\ba
   \tilde{\psi}(X) &:= \cf(\psi)(X) = \int_{G} \d g\ E_g(X)\, \psi(g) \,, \nn
   \psi(g) &= \cf^{-1}(\tilde{\psi})(g) = \int_{\fg^*} \frac{\d^d X}{(2\pi)^d}\ \overline{E_g(X)} \star_p \tilde{\psi}(X) \,, \nonumber
\ea
where $E_g(X)$, the kernel of the transform, is what we call the non-commutative plane wave. The explicit form of the non-commutative plane wave, and thus that of the transform, depends again on the choice of a quantization map or, equivalently, a deformation quantization $\star$-product. In fact, in terms of the canonical coordinates (of the first kind) $k(g) = -i\ln(g)\in\fg$ on $G$ obtained through the logarithm map, the plane wave is shown to be given by the star-exponential
\ba
E_g(X)=e_\star^{ik(g)\cdot X} \,, \nonumber
\ea
where $X\in \fg^*$.\footnote{We will use the physicists' convention of self-adjoint Lie algebra elements for unitary groups throughout.} In case $G$ has compact subgroups, the logarithm is multivalued, and we take $k(g)=-i\ln(g)$ to be in the principal branch. The introduced amended star-product $\star_p$ implements a projection onto the principal branch for the product of non-commutative plane waves. The set of plane waves $E_g(X)$ equipped with the $\star_p$-product then constitutes a representation of $G$, since $E_g(X) \star_p E_h(X) = E_{gh}(X)$. Hence, a given choice of quantization map uniquely determines the star-product and thus $E_g(X)$, which, in turn, uniquely determines the non-commutative Fourier transform and its inverse. This result also clarifies the relation with the so-called quantum group Fourier transform, extending again the work of Freidel \& Majid \cite{Freidel:2005ec}.

Last, we provide explicit examples of the above construction for three interesting choices of quantization maps: the symmetric map, the Duflo map, and the so-called Freidel-Livine-Majid map (used in the quantum gravity literature).\\

The outline of the paper is the following: in the next section \ref{sec:motivation} we motivate the general construction by working with the simplified case of Euclidean space, where the guiding ideas are easy to follow and the complications coming from the general Lie group structure are out of the way. Sections \ref{sec:algebra} and \ref{sec:groupfouriertransform} constitute the bulk of the article. We start by quantizing a Poisson subalgebra of the algebra of smooth functions on $T^*G$ as an abstract operator algebra $\fA$, emphasizing its underlying Hopf algebra structures inherited from the Lie group $G$ and Lie algebra $\fg^*$ structures. In Subsections \ref{ssec:position} and \ref{ssec:momentum} we define representations of $\fA$ in terms of functions on the group $G$ and the dual algebra $\fg^*$, respectively. And finally, in Section \ref{sec:groupfouriertransform} we derive the non-commutative plane wave that gives rise to the intertwiner between the aforementioned representations --- the non-commutative Fourier transform. Explicit examples in two distinctive cases, $\U(1)$ and $\SU(2)$, for various choices of quantization maps are worked out in the subsequent section, thus showing the existence of the algebra representation in some interesting cases. A short conclusion on the obtained results is given in Section \ref{sec:conc}.

\section{Motivation: Harmonic analysis on Euclidean space}
\label{sec:motivation}
To motivate the route we will follow next, let us understand the procedure for the simple case of Euclidean space, $\Rl^d$ ($d\in\Nl$), and see how the usual Fourier transform arises as an intertwiner between the position and momentum representations.\\

The classical phase space is given by $T^*\Rl^d=\Rl^d\times(\text{Lie}\,\Rl^d)^*$, where $(\text{Lie}\,\Rl^d)^*$ denotes the dual of the Lie algebra of $\Rl^d$, which coincides with $\Rl^d$ itself, $(\mathrm{Lie}\,\Rl^d)^* \cong (\Rl^d)^*\cong \Rl^d$. Let $\vec{x}=(x^i)$ and $\vec{p}=(p_j)$ ($i,j=1,\ldots,d$) be canonical coordinates in some basis on $\Rl^d$ and $(\text{Lie}\,\Rl^d)^*$, respectively, with Poisson brackets\footnote{Where appropriate, the equations should be read as holding for all values $i,j,k=1,\ldots, d$.}
\ba
\{x^i,x^j\}=0\,,\q\{x^i,p_j\}=\delta^i_j\,,\q \{p_i,p_j\}=0\,.
\label{eq:PbE}
\ea
The Poisson structure is defined directly on $C^{\infty}(T^*\Rl^d)$ by the canonical symplectic structure of the phase space and, together with the ordinary pointwise multiplication $\cdot$ on $C^\infty(T^*\Rl^d)$, gives rise to the full Poisson algebra $\cp_{\Rl^d}=(C^\infty(\Rl^{2d}),\{\cdot,\cdot\}, \cdot)$.\footnote{The pointwise product $\cdot$ is symmetric and associative, and $\{\cdot,\cdot\}$ is antisymmetric and satisfies the Jacobi identity. Furthermore, both structures are compatible in the sense that, for any $f,g,h\in C^\infty(T^*\Rl^d)$, $\{f,g\cdot h\}=\{f,g\}\cdot h+g\cdot \{f,h\}$, that is, the Leibniz rule `intertwines' pointwise multiplication and Poisson brackets.} As a physical system, we could think of $\cp_{\Rl^d}$ as the algebra of classical observables of a point particle moving on the Euclidean space, with $\vec{x}$ being the position, and $\vec{p}$ the respective canonical conjugate momentum.

We now seek to quantize this algebra $\cp_{\Rl^d}$, or a \emph{subalgebra} $\ca$ thereof, as an abstract operator $^*$-algebra $\fH$. That is, we want a map $\cq: \ca\rightarrow \fH$ such that the basic Poisson brackets (\ref{eq:PbE}) are mapped to the commutators
\ba
[X^i,X^j]=0\,,\q[X^i,P_j]=i\delta^i_j\1\,,\q [P_i,P_j]=0\,,
\label{eq:qPbE}
\ea
where $X^i=\cq(x^i)$, $P_j=\cq(p_j)$ are self-adjoint elements in $\fH$. The Lie algebra generated by $X^i$, $P_j$, and $\1$ is the usual Heisenberg algebra.\\

A few remarks about the map $\cq$ are in order:
\begin{itemize}
	\item $\cq(\ca) = \fH$ is, at this stage, an abstract operator $^*$-algebra. We may consider a representation of $\fH$ as a concrete operator algebra on a Hilbert space $\ch$, which is what we will do in the following. However, due to (\ref{eq:qPbE}), $X^i$ and $P_j$ are necessarily unbounded operators, and therefore their domains of definition have to be restricted to some dense subspaces of $\ch$ such that their images under the action of the operators are contained in $\ch$; or the treatment extended to a rigged Hilbert space \cite{rigged, nst}.
	\item $\cq$ is linear and satisfies $\cq(1)=\1$ and possibly $\cq(\phi(f))=\phi(\cq(f))$ for any function $\phi:\Rl\rightarrow \Rl$ for which $\cq(\phi(f)), \phi(\cq(f))$ are well defined (von Neumann rule).
	\item The need of a subalgebra $\ca\subset C^\infty(\Rl^{2d})$ comes from the general obstruction to quantizing consistently the full Poisson algebra $\cp_{\Rl^{d}}$, cf.\@ Groenewold-van Hove's theorem and generalizations thereof \cite{obstruction}. Even determining the maximal Lie subalgebra of $C^\infty(\Rl^{2d})$ for which quantization can be carried out is an open problem, and we again refer the reader to \cite{obstruction} for a detailed analysis of such subtleties. In the following, we shall be content with assuming the existence of such $\ca$, and will require it to be big enough to contain all the relevant functions of the subsequent analysis (in particular, exponentials). Moreover, it is also important that $\ca$ be complete in the sense that it guarantees local separation of points everywhere on the phase space.
\end{itemize}

As remarked above, we now consider representations $\pi$ of $\fH$ as a concrete algebra of (in general, unbounded) operators on some (dense subspace of a) Hilbert space $\ch$. In particular, $\pi: \fH\to \text{Aut}(\ch)$ is a linear $^*$-homomorphism between $\fH$ and the automorphisms of $\ch$, preserving commutators:
\ba
	\pi(\lambda A+\mu B)&=\lambda\pi(A)+\mu\pi(B) \,, \nonumber\\
	\pi(AB)&=\pi(A)\pi(B) \,, \nonumber\\
	\pi(A^*)&=\pi(A)^* \,, \nonumber\\
	\pi([A,B])&=[\pi(A),\pi(B)] \,, \nonumber
\ea
for all $A,B\in\fH$ and $\lambda, \mu\in \Rl$.\\

The commutativity of the $X^i$ operators allows to diagonalize all of them simultaneously. Accordingly, we have the \emph{position representation} $\pi_x$ of the algebra on $L^2(\Rl^d,\d^d x)$ on the joint spectrum of $X^i$'s such that
\ba
(\pi_x(X^i)\psi)(\vec{x})=x^i \psi(\vec{x})\,.
\ea
As already noted, the operators $X^i$ are unbounded, and therefore their domains must be restricted to a dense subset $C_c^\infty(\Rl^d) \subset L^2(\Rl^d,\d^d x)$ of smooth compactly supported functions on $\Rl^d$. Furthermore, since the operators $X^i$ constitute a maximal subset of commuting self-adjoint generators of the algebra $\fH$, the description of a state $\psi$ in $L^2(\Rl^d,\d^d x)$ is complete. To complete the description of the action of the operators, we note that by setting
\ba
(\pi_x(P_j)\psi)(\vec{x}) = -i\frac{\del}{\del x^j}\,\psi(\vec{x})\,,
\ea
we consistently represent the commutator $[X^i,P_j]=i\delta^i_j\1$, and thus this specification is shown to determine a representation of the original abstract operator $^*$-algebra $\fH$ on $L^2(\Rl^d,\d^d x)$. (The same remarks as before apply to the domains of $P_j$'s.) Anticipating our later considerations, we should note the important role the Leibniz rule of the partial derivatives with respect to the pointwise multiplication plays in reproducing the correct commutation relations. If one further requires irreducibility and regularity, this representation on $L^2(\Rl^d,\d^d x)$ is shown to be unique up to unitary equivalence due to the Stone-von Neumann theorem \cite{vN1,vN2}.\\

The same reasoning can be applied just as well, and independently, to the $P_j$'s. The diagonalization procedure gives another representation $\pi_p$ of $\fH$ on $L^2(\Rl^d,\d^d p/(2\pi)^d)$, where now the operators $P_j$ act multiplicatively
\ba
(\pi_p(P_j)\tilde{\psi})(\vec{p})= p_j\tilde{\psi}(\vec{p})\,.
\ea
Analogously, $\tilde{\psi}(\vec{p})$ are said to give a representation in terms of functions of the momenta, and $\pi_p$ is thus called a \textit{momentum representation}. Finally, the action of the operators $X^i$ in this basis which correctly reproduces the commutators $[X^i,P_j]=i\delta^i_j\1$ is given by
\ba
(\pi_p(X^i)\tilde{\psi})(\vec{p}) =i\frac{\del}{\del p_i}\,\tilde{\psi}(\vec{p})\,.
\ea
\\

We will now see that the usual Fourier transform $\cf$ is exactly the unique, unitary intertwiner between these two representations, a property we may write as $\pi_p(A)\circ\cf = \cf\circ\pi_x(A)$ for all $A\in\fH$, establishing, therefore, their equivalence. 

Hence, assuming that the two previous representations of $\fH$ are intertwined by an integral transform $\cf$, that is,
\ba
	\tilde{\psi}(\vec{p})\equiv \cf(\psi)(\vec{p}) := \int_{\Rl^d} \d^d x\, E(\vec{x},\vec{p})\, \psi(\vec{x})\,,\q \psi\in L^2(\Rl^d)\,, \nonumber
\ea
where $E(\vec{x},\vec{p})$ denotes the kernel of the transform, the intertwining property turns into properties for $E(\vec{x},\vec{p})$. On the one hand,
\ba
	(\pi_p(P_i) \cf(\psi))(\vec{p}) &= \int_{\Rl^d} \d^d x\, p_i\, E(\vec{x},\vec{p}) \psi(\vec{x})\,, \nn
	\cf(\pi_x(P_i) \psi)(\vec{p}) &= \int_{\Rl^d} \d^d x\, E(\vec{x},\vec{p})\, \left(-i\frac{\del}{\del x^i} \psi(\vec{x}) \right) = \int_{\Rl^d} \d^d x\, \left(i\frac{\del}{\del x^i} E(\vec{x},\vec{p}) \right)\, \psi(\vec{x})\,, \nonumber
\ea
where we used integration by parts for the last equality. (Note that smooth compactly supported functions vanish at infinity.) Therefore, for all $\psi\in L^2(\Rl^d)$ we have the differential equation
\ba\label{eq:Ep}
	p_i E(\vec{x},\vec{p}) = i\frac{\del}{\del x^i} E(\vec{x},\vec{p})\,.
\ea
On the other hand, from the corresponding requirement for the $X^i$ operators we get
\ba
	\cf(\pi_x(X^i)\psi)(\vec{p}) &= \int_{\Rl^d} \d^d x\, E(\vec{x},\vec{p})\, x^i \psi(\vec{x})\,, \nn
	(\pi_p(X^i) \cf(\psi))(\vec{p}) &= \int_{\Rl^d} \d^d x\, \left(i\frac{\del}{\del p_i} E(\vec{x},\vec{p}) \right)\, \psi(\vec{x})\,, \nonumber
\ea
which, for all $\psi\in L^2(\Rl^d)$, gives
\ba\label{eq:Ex}
	x^i E(\vec{x},\vec{p}) = i\frac{\del}{\del p_i} E(\vec{x},\vec{p})\,.
\ea
The unique and common solution to the two differential equations (\ref{eq:Ep}) and (\ref{eq:Ex}) is the plane wave $E(\vec{x},\vec{p}) = c\, e^{-i\vec{p}\cdot\vec{x}}$, where $c\in\Cl$ is an arbitrary integration constant. Hence, we find
\ba
	\tilde{\psi}(\vec{p}) \equiv \cf(\psi)(\vec{p}) = c\int_{\Rl^d} \d^d x\, e^{-i\vec{p}\cdot\vec{x}}\, \psi(\vec{x})\,.
\ea
For the particular value of $c=1$ the transform is found to be unitary, i.e., $\cf \circ \cf^* = \text{id}_{L^2(\Rl^d)} = \cf^* \circ \cf$ (and, in particular, invertible), the adjoint transform being given by
\ba
	\cf^{-1}(\tilde{\psi})(\vec{x}) = \int_{\Rl^d} \frac{\d^d p}{(2\pi)^{d}}\, \overline{E(\vec{x},\vec{p})}\, \tilde{\psi}(\vec{p}) = \psi(\vec{x}) \,.
\ea
Therefore, as advertized, $\pi_x$ and $\pi_p$ are unitarily equivalent with the Fourier transform $\cf$ their intertwiner.
\\

Let us further note an important property of the translations $(T_{\vec{y}}\psi)(\vec{x}) = \psi(\vec{x}+\vec{y})$. Since $\cf(T_{\vec{y}}\psi)(\vec{p}) = e^{i\vec{p}\cdot\vec{y}}\cf(\psi)(\vec{p})$, the translations act dually via pointwise multiplication by plane waves, and, therefore, the plane waves $e^{i\vec{p}\cdot\vec{x}}$ constitute a dual representation of the translation group. In fact, this follows directly from the form of the representations, since by integrating the action of partial derivatives we have $\psi(\vec{x}+\vec{y}) = e^{\vec{y}\cdot {\nabla}_{\vec{x}}}\psi(\vec{x}) = \pi_x(e^{i\vec{y}\cdot \vec{P}})\psi(\vec{x})$, and since $\cf$ intertwines the representations, $\cf(\pi_x(e^{i\vec{y}\cdot \vec{P}})\psi)(\vec{p}) = (\pi_p(e^{i\vec{y}\cdot \vec{P}})\cf(\psi))(\vec{p}) = e^{i\vec{y}\cdot \vec{p}}\tilde{\psi}(\vec{p})$. Notice, in particular, the important role that the global triviality of the Euclidean space plays here in integrating the action of the partial derivatives. Later, we will see that extra complications arise, if there are compact subgroups to the Lie group under consideration. These need to be properly taken care of in order for the translations to act dually by plane wave multiplication.\\

This derivation of the ordinary Fourier transform between the position and the momentum representations for $T^*\Rl^d$ motivates the line of thought that will be used in Sec.\ \ref{sec:algebra} for the general case of the cotangent bundle of a Lie group $T^*G$, and whose result, having first defined the two corresponding representations, will finally lead to the notion of non-commutative Fourier transform.

\section{Quantum representations for general (weakly exponential) Lie groups}
\label{sec:algebra}
We now turn to the case where the configuration space is a Lie group $G$ of the weakly exponential type, that is, such that the image of the exponential map, $\exp(\fg)\subset G$, is dense in $G$. The importance of this restriction will become clear, in particular, in Sec.\ \ref{sec:groupfouriertransform}, where one wants to be able to determine plane waves of the exponential type. Note that compact connected Lie groups are always exponential, since the exponential map commutes with conjugation and any compact connected Lie group is the union of the conjugates of a maximal torus, which is exponential.
A thorough summary of the status of the exponentiability of a Lie group and its complexity can be found in \cite{expstatus, explatest}.\\ 

The phase space of the system is given by the cotangent bundle $T^*G\cong G\times \fg^*$, which for Lie groups is always globally trivial, since we may always find a global basis of right (left) invariant covector fields through the pull-back of the multiplicative action of $G$ on itself $R_h:G\rightarrow G$, $g\mt gh$ ($L_h:G\rightarrow G$, $g\mt hg$), $h\in G$.
Cotangent bundles are endowed with a canonical symplectic structure that, together with ordinary pointwise multiplication $\cdot$ on $C^\infty(T^*G)$, uniquely determines the Poisson algebra $\cp_G=(C^\infty(T^*G),\{\cdot,\cdot\}, \cdot)$,\footnote{The canonical symplectic 1-form $\theta$ on $T^*G$ is obtained via the pull-back $\pi^*: T^*G \rightarrow T^*(T^*G)$ of the canonical bundle projection $\pi: T^*G \rightarrow G$, $\pi(\alpha) = p \in G$ for all $\alpha \in T_p^*G$. The symplectic 2-form is then obtained as $\omega = -\d\theta$. To any $f \in C^\infty(T^*G)$ can then be associated a vector field $X_f$ on $T^*G$ via the relation $\omega(X_f,\cdot) = \d f$. The Poisson bracket for functions $f,g \in C^\infty(T^*G)$ is then given canonically by $\{f,g\} := \omega(X_f,X_g) \in  C^\infty(T^*G)$ \cite{Vilasi:2001bm}.} and for any functions $f,g \in C^\infty(T^*G)$ we obtain
\ba
	\{f,g\} \equiv \frac{\del f}{\del X_i} \cl_ig - \cl_if \frac{\del g}{\del X_i} + c_{ij}^{\phantom{ij}k} \frac{\del f}{\del X_i} \frac{\del g}{\del X_j} X_k\,,
	\label{eq:generalPB}
\ea
where $\cl_i$ are Lie derivatives on $G$ with respect to an orthonormal basis of right-invariant vector fields, $X_i$ are Euclidean coordinates on $\fg^* \cong \Rl^d$, $d:=\dim(G)$, $c_{ij}^{\phantom{ij}k}$ the structure constants of the Lie algebra $\fg$ ($\cong \fg^*$), $i,j,k=1,\ldots, d$, and Einstein summation convention is assumed. 

We now seek to quantize this algebra, or at least a \emph{maximal subalgebra} $\ca$ thereof for which this is consistent, as an abstract operator $^*$-algebra $\fA$. We define a quantization map $\cq: \ca\rightarrow \fA$ such that $\cq(f)=: \hat{f}$ for all $f \in \ca_G\subset C^\infty(G)$, and $\cq(X_j)=:\hat{X}_j$, satisfying
\ba
  [\hat{f},\hat{g}] = 0\,,\quad [\hat{X}_i,\hat{f}] = i\widehat{{\cl}_i f} \in \fA_G\,,\quad [\hat{X}_i,\hat{X}_j] = i c_{ij}^{\phantom{ij}k}\hat{X}_k\,, \label{eq:comm}
\ea
for all $\hat{f},\hat{g}\in \fA_G$. We denoted by $\ca_G$ the subalgebra of $\ca\subset C^\infty(G\times \fg^*)$ of functions constant in the second argument, and $\fA_G := \cq(\ca_G)$, which is a commutative subalgebra of $\fA$.

In general, we cannot introduce differentiable coordinates $\zeta^i\in C^\infty(G)$ on $G$ due to a global obstruction, in particular, if $G$ has compact subgroups. Accordingly, we cannot have operators in $\fA$ corresponding to coordinates on $G$. However, such coordinates can be approximated arbitrarily well by elements in $C^\infty(G)$, and we may define coordinate operators $\hat{\zeta}^i$, not necessarily in $\fA_G$, corresponding to a set of coordinates $\zeta^i: G \rightarrow \Rl$ by imposing $\hat{f} \stackrel{!}{=} f_\zeta(\hat{\zeta}^i)$, where $f_\zeta \circ \vec{\zeta} \equiv f$, for all $f\in C^\infty(G)$. We then have formally the commutators
\ba
  [\hat{\zeta}^i,\hat{\zeta}^j] = 0 \,,\quad [\hat{X}_i,\hat{\zeta}^j] = i\widehat{{\cl}_i \zeta^j} \,,\quad [\hat{X}_i,\hat{X}_j] = i c_{ij}^{\phantom{ij}k}\hat{X}_k\,. \label{eq:commzeta}
\ea
Further assuming that $\zeta^i(e)=0$ and $\cl_i\zeta^j(e)=\delta_i^j$, the explicit form of the operator $\widehat{{\cl}_i\zeta^j}$ may be obtained (in a neighborhood of the identity) from the Taylor series expansion of the Lie derivatives at the identity in terms of the coordinates
\ba
	\cl_i\zeta^j(g) = \sum_{n=1}^{\infty} C^j_{i q_1 \cdots q_{n-1}} \zeta^{q_1}(g) \cdots \zeta^{q_{n-1}}(g)\,, \nonumber
\ea
simply as
\ba\label{eq:Lzeta}
	\widehat{{\mc{L}}_i\zeta^j} = \sum_{n=1}^{\infty} C^j_{i q_1 \cdots q_{n-1}} \hat{\zeta}^{q_1} \cdots \hat{\zeta}^{q_{n-1}}\,,
\ea
where $C^j_{i q_1 \cdots q_{n-1}}\in\Rl$ are constant coefficients specific to the chosen coordinates. Clearly, we are always free to change coordinates as $\fA_G$ is commutative. The same remarks for the quantization map $\cq$ on $\cp_{\Rl^d}$ apply {\it ipsis verbis} with $\Rl^d$ replaced by $G$. 

We will call the algebra generated by $\hat{f} \in \fA_G$ and $\hat{X}_i$, already denoted by $\fA$, as the \emph{quantum algebra} for $T^*G$. Note that it may differ from the Heisenberg algebra $\fH$ as now the commutator $[\hat{X}_i,\hat{\zeta}^j]$ does not in general equal a multiple of $\1$ for any choice of coordinates $\zeta^j$.\\

The quantum algebra $\fA$ has, in fact, some extra structure inherited from the Lie group and Lie algebra structures of $G$ and $\fg$. On the one hand, notice that the commutation relations for the $\hat{X}_i$ operators among themselves coincide with the Lie algebra commutation relations for $\fg$. Therefore, the restriction of $\cq$ onto functions $\ca_{\fg^*}\subset C^\infty(\fg^*) \subset C^\infty(G\times\fg^*)$ that are constant in the first factor maps to the completion of the universal enveloping algebra of $\fg$, $\fA_{\fg^*} := \cq(\ca_{\fg^*}) \cong \overline{U(\fg)} \subset \fA$. $U(\fg)$ is endowed with a natural Hopf algebra structure with coproduct $\Delta_{\fg^*}$, counit $\eps_{\fg^*}$, and antipode $S_{\fg^*}$, which extends to a corresponding structure on $\fA_{\fg^*}$ given by
\ba
\label{eq:hopfXD}
\Delta_{\fg^*} & : \fA_{\fg^*} \rightarrow \fA_{\fg^*}\otimes \fA_{\fg^*}\,,\hspace{30pt} \Delta_{\fg^*}( \hat{X}_i ) = \hat{X}_i\otimes \1 + \1 \otimes \hat{X}_i\,,\\
\label{eq:hopfXe}
\eps_{\fg^*} & : \fA_{\fg^*} \rightarrow \Rl \,,\hspace{78pt} \eps_{\fg^*}(\1) = 1,\hspace{18pt}  \eps_{\fg^*}(\hat{X}_i)  = 0\,,\\
\label{eq:hopfXS}
S_{\fg^*} & : \fA_{\fg^*} \rightarrow \fA_{\fg^*} \,,\hspace{67pt} S_{\fg^*}(\1) = \1,\hspace{15pt} S_{\fg^*}(\hat{X}_i) = -\hat{X}_i\,.
\ea

On the other hand, the structure maps of $G$, that is, the group multiplication $G \times G \rightarrow G\,, (g,h)\mapsto gh$, the inclusion of the unit $\{e\} \hookrightarrow G\,, e\mapsto e$, and the inversion map $G\rightarrow G\,, g\mapsto g^{-1}$, induce, respectively, the following algebra homomorphisms on $C^\infty(G)$,
\bas
\Delta & : C^\infty(G) \rightarrow C^\infty(G\times G)\,,\hspace{40pt} \Delta (f)(g,h) = f(gh)\,,\\
\eps & : C^\infty(G) \rightarrow \Rl\,,\hspace{112pt} \eps(f)  = f(e)\,,\\
S & : C^\infty(G) \rightarrow C^\infty(G)\,,\hspace{72pt} S(f)(g) = f(g^{-1})\,.
\eas
Equipped with these structure maps, $C^\infty(G)$ forms nearly a Hopf algebra.\footnote{The problem is that the target of the map $\Delta$ is $C^\infty(G\times G)$ and not the algebraic tensor product $C^\infty(G)\otimes C^\infty(G)$. We can identify $C^\infty(G)\otimes C^\infty(G)$ with a subspace of $C^\infty(G\times G)$, but the image of $\Delta$ is not contained in this subspace unless $G$ is finite. However, each unital subalgebra $\mathfrak{a}\subseteq C^\infty(G)$ which satisfies $\Delta(\mathfrak{a})\subseteq \mathfrak{a}\otimes \mathfrak{a}$ and $S(\mathfrak{a})\subseteq \mathfrak{a}$ is a Hopf algebra with respect to the restriction of the maps $\Delta, \eps$ and $S$.} To obtain the corresponding Hopf algebra structure in $\fA_G$ for any exponential Lie group, consider the canonical coordinates (of the first kind) $k:G\rightarrow \fg \cong \Rl^d$, $g \mapsto -i\ln(g)$ obtained through the logarithm map. As these coordinates satisfy $k(e)=0$ and $k(g^{-1})=-k(g)$, by correspondence to the above structure, we may set for the corresponding operators $\eps_G(\hat{k}^i) = 0$ and $S_G(\hat{k}^i) = -\hat{k}^i$. Furthermore, we may write
\ba
	k^i(gh) = & \sum_{n=1}^{\infty} \sum_{\substack{k,l\in\Nl\\k+l=n}} B^i_{p_1\cdots p_k q_1\cdots q_l} k^{p_1}(g) \cdots k^{p_k}(g) k^{q_1}(h) \cdots k^{q_l}(h) \,, \label{eq:kexpand}
\ea
where $B^i_{p_1\cdots p_k q_1\cdots q_l} \in \Rl$ are constant coefficients. This is just the Baker-Campbell-Hausdorff formula for $G$, denoted in the following by $k^i(gh) \equiv \cb(k(g),k(h))^i$. In the lowest order in $|k|$ we have $k^i(gh) \approx k^i(g) + k^i(h)$, and the higher orders encode the non-linearity of the group manifold. Notice that, if the logarithm for $G$ is multivalued --- which is the case if $G$ has compact subgroups ---, in general, the result $k(gh)$ does not lie in the principal branch of the logarithm even if $k(g)$ and $k(h)$ do. We may then define the coproduct for the corresponding coordinate operators as
\ba
	\Delta_G(\hat{k}^{i}) = & \sum_{n=1}^{\infty} \sum_{\substack{k,l\in\Nl\\k+l=n}} B^i_{p_1\cdots p_k q_1\cdots q_l}\hat{k}^{p_1} \cdots \hat{k}^{p_k} \otimes \hat{k}^{q_1} \cdots \hat{k}^{q_l} \,,
\ea
which reflects the group structure. The coproduct corresponding to that of $f\in \ca_G$ in $\fA_G$ can then be formally defined as
\bas
	\Delta_G(\hat{f}) \equiv f_k(\Delta_G(\hat{k}^i)) \,,
\eas
where $f_k: \fg \cong \Rl^d \rightarrow \Cl$ is the lift of $f:G\rightarrow \Cl$ onto the Lie algebra as $f_k(k) \equiv f(e^{ik})$. Clearly, by this definition of the coproduct, the possible multivaluedness of $k$ is taken care of by the corresponding periodicity in $f_k$. The explicit meaning of this rather formal expression can be understood locally (for analytic functions) by expanding $f_k$ as a power series in $k^i$.

Similarly, we can consider parametrizations $\zeta:G \rightarrow \fg \cong \Rl^d$ of $G$ other than the canonical coordinates. Given $\zeta_k^j(\vec{k}(g))$, we may write accordingly
\ba
	&\Delta_G(\hat{\zeta}^{i}) = \zeta_k^i(\Delta_G(\hat{k}^i)) =\sum_{n=1}^{\infty} \sum_{\substack{k,l\in\Nl\\k+l=n}} C^i_{p_1\cdots p_k q_1\cdots q_l} \hat{\zeta}^{p_1} \cdots \hat{\zeta}^{p_k} \otimes \hat{\zeta}^{q_1} \cdots \hat{\zeta}^{q_l} \,, \label{eq:zetacoprod}
\ea
where the new coefficients $C^i_{p_1\cdots p_k q_1\cdots q_l}\in\Rl$ are obtained from the expression (\ref{eq:kexpand}) for $k$-coordinates by expanding $\zeta^i(g)$ in $k^i(g)$. Notice that the coefficients $C^i_{p_1\cdots p_k q_1\cdots q_l}$ here are the same as those appearing in (\ref{eq:Lzeta}) for coordinates such that $\zeta_k^i(\vec{0})=0$ and $\frac{\prt}{\prt k^i}\zeta_k^j(\vec{0})=\delta_i^j$. This will be important in reproducing correctly the commutators in the algebra representation defined below.

The significance of these Hopf structures cannot be underestimated, in particular, with respect to the coproducts $\Delta_{\fg^*}$ and $\Delta_G$, and how they ensure the correct reproduction of the commutation relations in the two representations of $\fA$ we now proceed to define.\\

We now turn to explicit representations $\pi$ of the quantum algebra $\fA$ as a concrete operator algebra on some Hilbert space $\ch$, where, as before, $\pi:\fA\to \text{Aut}(\ch)$ is a linear $^*$-homomorphism preserving commutators.

\subsection{Group representation $\pi_G$}
\label{ssec:position}
The \emph{group representation} $\pi_G$ on $L^2(G)$ is defined as the one diagonalizing all the operators $\hat{f}\in\fA_G$:
\ba
	(\pi_G(\hat{f})\psi)(g) \equiv f(g)\psi(g)\,,
\label{eq:rep1}
\ea
for all $f\in\ca_G$ such that $\hat{f} \equiv \cq(f)$, as before. The resulting function $f\psi$ will not in general lie in $L^2(G)$ for all $\psi\in L^2(G)$, but we may again restrict the domain of $\pi_G(\hat{f})$ to be the subspace of $\ca_G$ of smooth compactly supported functions $C^\infty_c(G)$ on $G$ --- dense in $L^2(G)$ ---, so that $f\psi\in C^\infty_c(G)$ for all $\psi\in C^\infty_c(G)$. For the Lie algebra operators $\hat{X}_i$ we may set
\ba
	(\pi_G(\hat{X}_i) \psi)(g) \equiv i{\cl}_i\psi(g)\,,
\label{eq:rep2}
\ea
where ${\cl}_i$ are again the Lie derivatives with respect to an orthonormal basis of right-invariant vector fields on $G$, and similar remarks as above hold about the domain of $\pi_G(\hat{X}_i)$. One can easily check that the commutation relations (\ref{eq:comm}) are correctly reproduced, so that the above actions define a representation of $\fA$. As usual, the inner product is given for $\psi,\psi'\in L^2(G)$ by
\ba
	\langle \psi,\psi' \rangle_G \equiv \int_G \d g\ \overline{\psi(g)}\,\psi'(g) \,,
\ea
where $\d g$ is the right-invariant Haar measure on $G$.

To prove that (\ref{eq:rep1}), (\ref{eq:rep2}) give a representation of (\ref{eq:comm}) we used, in fact, a fairly innocent property of the Lie derivative: $\cl_i$ satisfies the \emph{usual} Leibniz rule with respect to the pointwise product of functions, that is, $\cl_i(ff')=(\cl_if)f'+f(\cl_if')$. Even though we know this to be true by other means, this can be expressed as a compatibility condition between the coproduct $\Delta_{\fg^*}$ of $\fA_{\fg^*}$ and the pointwise product $m_G: f \otimes f' \mapsto f\cdot f'$ for $f,f'\in C^\infty(G)$, namely, 
\ba
\pi_G(\hat{X}_i) \circ m_G = m_G\circ (\pi_G\otimes \pi_G)(\Delta_{\fg^*}(\hat{X}_i))\,,
\label{eq:leibnizG}
\ea
where $\pi_G\otimes\pi_G$ denotes the tensor product of the representation $\pi_G$. More simply, (\ref{eq:leibnizG}) amounts to $\cl_i \circ m_G = m_G \circ \Delta_{\fg^*}(\cl_i)$, which on a tensor product $f\otimes f'$ gives
\ba
\cl_i(f\cdot f') &=\cl_i\circ m_G(f\otimes f')  \nn &= m_G \circ \Delta_{\fg^*}(\cl_i)(f\otimes f')  \nn &= m_G (\cl_i f \otimes f' + f\otimes \cl_i f') \nn &= (\cl_i f)\cdot f'+ f\cdot(\cl_if')\,, \nonumber
\ea
that is, the usual Leibniz rule for the pointwise product. Notice that while the Leibniz rule is a representation-dependent concept, the coproduct is representation-independent. Essentially, (\ref{eq:leibnizG}) can be seen as consistency of the representation of the operator $\pi_G(\hat{X}_i)$ and the pointwise multiplication, with the underlying Hopf algebra structure of $\fA$. Different elements in the given representation will have, in principle, different multiplications such that the compatibility with the Hopf algebra structure (in particular, the coproduct) of $\fA$ is satisfied. For instance, the analogous expression for $\hat{\zeta}^i$ is $\pi_G(\hat{\zeta}^i)\circ m_* = m_*\circ (\pi_G\otimes\pi_G)(\Delta_G(\hat{\zeta}^i))$, which is satisfied for the convolution product $m_*$.\\

Since it will be crucial for defining the \emph{algebra representation}, let us state this requirement more generally. Let $\pi$ be representation of $\fA$ on a space $\fF_m$ with $m: f \otimes f' \mapsto f \cdot_m f'$ the corresponding multiplication. The compatibility with the coproduct $\Delta$ can be written in an abstract form as the identity $\pi(\hat{T}) \circ m = m \circ (\pi\otimes\pi)(\Delta(\hat{T}))$, for $\hat{T}$ an operator in $\fA$. That is, the following diagram
\ba
  \begin{CD}
    \fF_m \otimes \fF_m @>{m}>> \fF_m  \\
    @V{\pi\otimes\pi(\Delta(\hat{T}))}VV            @VV{\pi(\hat{T})}V\\
    \fF_m \otimes \fF_m @>{m}>> \fF_m
  \end{CD}
\label{eq:leibnizCD}
\ea
commutes. It is clear that the diagram does not commute for all products and coproducts. However, given a coproduct, it tells which product makes it commute for the chosen operator $\hat{T}$ in the given representation and, therefore, compatible with the Hopf algebra structure in the sense of the diagram. Equivalently, reverting the logic, given a product and a coproduct, (\ref{eq:leibnizCD}) tells how a certain representation of an operator $\hat{T}$ acts on an $m$-product of functions, i.e., a generalized Leibniz rule for $\pi(\hat{T})$.

\subsection{Algebra representation $\pi_{\fg^*}$}
\label{ssec:momentum}
We would now like to have a representation naturally acting on functions of the classical dual space $\fg^*$, according to the decomposition of the phase space $T^*G \cong G\times\fg^*$. That is, functions $\varphi(X)$ analogous to functions of the classical coordinates on $\fg^*$.

However, the route taken to obtain the group representation, based on simultaneous diagonalization of the operators $\hat{f} \in \fA_G$ can no longer be used because $\hat{X}_i\in\fA_{\fg^*}$ are non-commuting.
In other words, since the action $(\pi_{\fg^*}(\hat{X}_i) \varphi)(X)=X_i\varphi(X)$ cannot possibly make sense in general, due to the non-zero Lie algebra structure constants $c_{ij}^{\ \ k}$, we introduce an operation that suitably \emph{deforms} it, giving the needed freedom to satisfy the commutation relations. We will denote it by a star-product $\star$, and define for all $i=1,\ldots, d$
\ba
	(\pi_{\fg^*}(\hat{X}_i) \varphi)(X) := X_i \star \varphi(X)\,. 
\ea
Notice that the commutator $[\hat{X}_i,\hat{X}_j] = i c_{ij}^{\phantom{ij}k}\hat{X}_k$ turns into
\ba
 (X_i\star X_j-X_j\star X_i) \star \varphi(X) = ic_{ij}^{\phantom{ij}k} X_k \star \varphi(X)\,, \nonumber
\ea
giving a condition on the $\star$-product. In fact, we will impose the stronger condition
\ba
	(\pi_{\fg^*}(f(\hat{X}_i)) \varphi)(X) = f_\star(X) \star \varphi(X)\,,
\ea
for all $f_\star\in \ca_{\fg^*}\subset C^\infty(\fg^*)$ such that $f(\hat{X}_i) = \cq(f_\star) \in \fA_{\fg^*}$. This guarantees that $f_\star$ has the interpretation of the function which upon quantization gives $f(\hat{X}_i)$, and so establishes a connection between the classical phase space structure and the quantum operators. We then have
\ba
	(\pi_{\fg^*}(\cq(f_\star)\cq(f'_\star)) \varphi)(X) &= (\pi_{\fg^*}(f(\hat{X}_i))\pi_{\fg^*}(f'(\hat{X}_i)) \varphi)(X) \nn
	&= f_\star(X) \star f'_\star(X) \star \varphi(X) \nn
	&= (\pi_{\fg^*}(\cq(f_\star \star f'_\star)) \varphi)(X) \nonumber
\ea
for all $f_\star,f'_\star\in \ca_{\fg^*}$. 
Therefore, the $\star$-product and the quantization map $\cq$ are related by
\ba
f_\star \star f'_\star = \cq^{-1}(\cq(f_\star)\cq(f'_\star))\,,
\label{eq:definitionstar}
\ea
which is the idea of star-products defined in the context of deformation quantization \cite{defquant}.\footnote{Associativity and $1\star f_\star =f_\star= f_\star\star 1$, $f_\star\star f'_\star-f'_\star\star f_\star= i\{f_\star,f'_\star\}$ are easily verified using the properties of $\cq$.}

In other words, \emph{the choice of quantization map determines uniquely the $\star$-product to be used in representing the quantum algebra in terms of functions on $\fg^*$}.

We note that in order for $\fA_{\fg^*} \equiv \cq(\ca_{\fg^*})$ to be closed under operator product, a $\star$-product of functions on $\ca_{\fg^*}$ must again lie in $\ca_{\fg^*}$. This imposes some natural continuity and convergence requirements on the $\star$-product, which we assume to be fulfilled in the following.\\

Before moving on to define the algebra representation, and identifying the properties that the $\star$-product has to satisfy for this to exist, let us give a few more details on the properties of quantization maps, and of the resulting $\star$-products.
 
As remarked before, the image of the quantization map restricted to functions constant in the first factor, that is, $\fA_{\fg^*}:=\cq(\ca_{\fg^*})$, \emph{amounts to a completion of the universal enveloping algebra} $U(\fg)$. Of course, $\ca_{\fg^*}\subset C^\infty(\fg^*)$ may be too big a space, and we can make do with the space of polynomials in $\fg^*$, $\text{Pol}(\fg^*)$, which is known to be (graded) isomorphic to the symmetric algebra $\text{Sym}(\fg)$ of $\fg$. The Poincar\'e-Birkhoff-Witt theorem then states that the latter is isomorphic to the universal enveloping algebra $U(\fg)$ (as a filtered vector space). The important point is that $U(\fg)$ \emph{can be identified with the algebra of right-invariant differential operators on $G$}, the natural ground for the algebra of a quantum theory. (See Appendix \ref{app:uea} for more details.) Further, \emph{the quantization map $\cq$, when restricted to $\text{Sym}(\fg)$, provides an isomorphism and, in particular, encodes the operator ordering ambiguity coming from the non-commutativity of the elements $\hat{X}_i\in U(\fg)$}. For example, we could choose \emph{standard ordering} $\cq(X_i^nX_j^m)=\hat{X}_i^n\hat{X}_j^m$, or \emph{Weyl ordering} $\cq(X_i^nX_j^m)=\cs(X_i^nX_j^m)$ where $\cs$ is the total symmetrization map (\ref{eq:symmetrizationmap}), or ordering coming from the \emph{Duflo map} $\cd$ (\ref{eq:duflomap}) $\cq(X_i^nX_j^m)=\cd(X_i^nX_j^m)$, all depending on the properties we want to preserve. The star-product on $\text{Pol}(\fg^*)$ inherits these same properties, as it is constructed from the non-commutative product of the differential operators exactly in order to mimic their behavior. More generally, the star-product can be written as a formal power series with expansion parameter $\hbar$:
\ba\label{eq:starprodexpansion}
f_\star\star f'_\star= f_\star f'_\star + \sum_{k=1}^\infty\hbar^kB_k(f_\star,f'_\star)\,,
\ea
where $B_k$ are linear bidifferential operators of degree at most $k$, making quantization as a deformation of the commutative pointwise product explicit. In general, this series diverges, and convergence has to be established for suitable subalgebras.

Notice, however, also that for the completion $\overline{U(\fg)}$ the one-to-one correspondence with right-invariant differential operators may be partially lost. In particular, if exponentials $e^{ik}$, $k\in\fg$, belong to the completion, and $G$ has compact subgroups, there are $k(e) \neq 0$ in $\fg$ such that $e^{k(e)\cdot\vec{\cl}} = 1$. These are the branched values of the logarithm $k(e)=-i\ln(e)$, where $e\in G$ denotes the identity element. The set of elements $\ci := \{e^{ik}\in\overline{U(\fg)}:k=-i\ln(e)\}$ forms a multiplicative normal subgroup of $\overline{U(\fg)}$ and it is then natural to consider the elements of $\overline{U(\fg)}$ modulo $\ci$ to restore the one-to-one correspondence. We will come back to this important point in the next section.\\

Now, let $\star$ be a deformation quantization star-product for $U(\fg)$, extended to $\fA_{\fg^*}$, and let $\hat{\zeta}^i$ be (coordinate) operators corresponding to a specific parametrization of $G$, as defined in the beginning of this section. We \emph{define} the representation of the operators $\hat{\zeta}^i$ and $\hat{X}_i$ acting on the space of smooth compactly supported functions $\varphi\in C^\infty_c(\fg^*)$ on $\fg^*$ to be
\ba
  (\pi_{\fg^*}(\hat{X}_i) \varphi)(X) & \equiv X_i \star \varphi(X)\,,\nn 
  (\pi_{\fg^*}(\hat{\zeta}^i) \varphi)(X) & \equiv -i\del^i \varphi(X) \,,
\label{eq:momentumrep}
\ea
where we denote $\del^i := \frac{\del}{\del X_i}$, and by the second equation we explicitly mean
\ba
	(\pi_{\fg^*}(\hat{f}) \varphi)(X) \equiv f_k(-i\vec{\del}) \varphi(X) \,, \nonumber
\ea
where $f_k(k) := f(e^{ik}) \in C^\infty(\fg)$ for all $f\in C^\infty(G)$. It is clear from the power series expansion (\ref{eq:starprodexpansion}) of the $\star$-product that the result of these actions is again compactly supported, and therefore $C^\infty_c(\fg^*)$ is closed under these actions.

Now we proceed to identify the properties that the $\star$-product has to satisfy in order for the above equations to define a faithful representation of the fundamental quantum algebra $\fA$. Due to the properties of the deformation quantization $\star$-product, the first equation in (\ref{eq:momentumrep}) guarantees, by construction, that the observables depending only on $\hat{X}_i$ (up to finite order) are represented through an algebra isomorphism. Similarly, since the partial derivative operators on $\fg^*$ are commutative, $\hat{f}\mapsto\pi_{\fg^*}(\hat{f}),\ \hat{f}\in\fA_{\fg^*}$, is clearly a homomorphism. 
Therefore, in order to show that we have a representation of the quantum algebra, the only non-trivial part is to show that the commutator $[\hat{X}_i,\hat{\zeta}^j]$ is correctly reproduced, namely, due to (\ref{eq:Lzeta}) we should find
\ba
	& (\pi_{\fg^*}([\hat{X}_i,\hat{\zeta}^j]) \varphi)(X) = i\sum_{n=1}^\infty C^j_{iq_1\cdots q_{n-1}} (\pi_{\fg^*}(\hat{\zeta}^{q_1}) \cdots \pi_{\fg^*}(\hat{\zeta}^{q_{n-1}}) \varphi)(X) \,. \label{eq:Xcomm}
\ea
Now, the left-hand-side reads
\ba
	\pi_{\fg^*}([\hat{X}_i,\hat{\zeta}^j]) \varphi &= [\pi_{\fg^*}(\hat{X}_i),\pi_{\fg^*}(\hat{\zeta}^j)] \varphi \nn
	&= -i X_i \star (\del^j\varphi) + i\del^j(X_i \star \varphi) \,. \nonumber
\ea
In order to compute the second term, we must know how the partial derivative acts on $\star$-products of functions. Here, we will again impose the compatibility of the coproduct of the operator algebra and the algebra multiplication, expressed neatly by the commutative diagram (\ref{eq:leibnizCD}). In other words, we require that
\ba\label{eq:leibnizg}
	\pi_{\fg^*}(\hat{\zeta}^i) \circ m_{\fg^*} = m_{\fg^*} \circ (\pi_{\fg^*}\otimes\pi_{\fg^*})(\Delta_{G}(\hat{\zeta}^i)) \,,
\ea
where $m_{\fg^*}: f\otimes f \mapsto f\star f'$. Explicitly, using the coproduct formula (\ref{eq:zetacoprod}), imposing this requirement gives
\ba
& (-i{\del^i})(f \star f') = \sum_{n=1}^{\infty} \sum_{\substack{k,l \in \Nl \\k+l=n}} C^i_{p_1\cdots p_k q_1\cdots q_l} \left[\left(-i{\del^{p_1}}\right) \cdots \left(-i{\del^{p_k}}\right) f\right] \star \left[\left(-i{\del^{q_1}}\right) \cdots \left(-i{\del^{q_l}}\right) f'\right] \,, \nonumber
\ea
and thus we obtain
\ba
	\prt^j(X_i \star \varphi) = &\ X_i \star C^j_k(\del^k\varphi) + \sum_{n=1}^{\infty} C^j_{i q_1 \cdots q_{n-1}} ((-i\del^{q_1}) \cdots (-i\del^{q_{n-1}})\varphi) \,. \nonumber
\ea
Assuming $C_i^j \equiv \frac{\del}{\del k^i}\zeta_k^j(0)=\delta_i^j$ at the origin of the coordinates, we have then
\ba
	&-i X_i \star (\del^j\varphi) + i\del^j(X_i \star \varphi) = i \sum_{n=1}^{\infty} C^j_{i q_1 \cdots q_{n-1}} ((-i\del^{q_1}) \cdots (-i\del^{q_{n-1}})\varphi) \,, \nonumber
\ea
which is exactly the right-hand side of (\ref{eq:Xcomm}).
Therefore, \emph{if the $\star$-product satisfies the property encoded in the commutative diagram} (\ref{eq:leibnizCD}), then the commutator is correctly reproduced through the action (\ref{eq:momentumrep}) of the operators, and therefore $\pi_{\fg^*}$ defines a representation of $\fA$ in terms of a specific choice of coordinates on the group used in defining $\fA$ itself. In fact, the compatibility condition can also be interpreted as a condition between the choice of quantization map, thus of $\star$-product, and the choice of coordinates on the group.

\

Let us recapitulate what we have shown for the algebra representation thus far. Assume that
\begin{itemize}
	\item[(i)] $\fA_{\fg^*} := \cq(\ca_{\fg^*})$ is a subalgebra of the full quantum algebra $\fA$, where $\ca_{\fg^*}\subset C^\infty(\fg^*)$,
	\item[(ii)] the coproduct $\Delta_{G}$ is compatible with the operator product in $\fA_{\fg^*}$, Equation (\ref{eq:leibnizg}), in the sense of the commutative diagram (\ref{eq:leibnizCD}), and
	\item[(iii)] coordinates $\zeta: G \rightarrow \fg \cong \Rl^d$ on $G$ satisfy $\zeta_k^i(0)=0$ and $\frac{\prt}{\prt k^i} \zeta_k^j(0)=\delta_i^j$ for all $i,j=1,\ldots,d$, where $\zeta_k(k) \equiv \zeta(e^{ik})$.
\end{itemize}
Then, the action of the operators in (\ref{eq:momentumrep}),
\ba
  (\pi_{\fg^*}(\hat{X}_i) \varphi)(X) &\equiv X_i \star \varphi(X) \,, \nn
  (\pi_{\fg^*}(\hat{\zeta}^i) \varphi)(X) &\equiv -i\del^i \varphi(X) \nonumber
\ea
defines a representation of $\fA$ on $C^\infty_c(\fg^*)\ni \varphi$, which we call the \emph{algebra representation} $\pi_{\fg^*}$.

We remark once more that we have not provided a constructive definition, and that the existence of the algebra representation for a given quantization map and $\star$-product is not guaranteed \emph{a priori}. Instead, we have identified the properties that such $\star$-product has to satisfy for the representation to exist, to be checked for each given choice of quantization map. It is clear that, in general, that is, for arbitrary quantization map and $\star$-product, these requirements need not be satisfied, and no algebra representation thus exists. On the other hand, we show in the following that these properties are in fact fulfilled for various interesting choices of quantization maps, so the construction is at the same time non-trivial and non-empty.

\

Finally, with the above assumption (i) implying that a $\star$-product of functions in $C_c^\infty(\fg^*)$ for the deformation quantization corresponding to $\cq$ is again in $C_c^\infty(\fg^*)$, we have the sesquilinear form for $\varphi,\varphi'\in C_c^\infty(\fg^*)$ given by
\ba
	\langle \varphi,\varphi' \rangle_{\fg^*} := \int_{\fg^*} \frac{\d^d X}{(2\pi)^d}\ (\overline{\varphi} \star \varphi')(X) \,.
\ea
This form is, in general, degenerate, i.e., the set of functions $\cn := \{\varphi\in C_c^\infty(\fg^*): \langle \varphi,\varphi \rangle_{\fg^*}=0 \}$ may be non-empty. To define a proper inner product and the corresponding norm completion, which would then be our Hilbert space, we should quotient $C_c^\infty(\fg^*)$ by the degenerate subspace $\cn$. Furthermore, to be consistent with the action of $\fA$, we should also show that $\cn$ is invariant under that action. The latter is the non-trivial part, and for the time being, we will simply assume that this can be done, and denote the completion of $C_c^\infty(\fg^*)/\cn$ in the norm $\norm{\varphi} \equiv \sqrt{\langle\varphi,\varphi\rangle_{\fg^*}}$ as $L^2_\star(\fg^*)$. The existence of a unitary intertwiner between the two representation spaces $L^2(G)$ and $L^2_\star(\fg^*)$, which will be shown in the next section, will eventually justify this assumption.

\section{The non-commutative Fourier transform}
\label{sec:groupfouriertransform}
Our next objective is to find the relation between the two representations $\pi_G$ and $\pi_{\fg^*}$ of $\fA$ defined above. In correspondence with the Euclidean case presented in the Motivation section \ref{sec:motivation}, we will assume that there exists an intertwiner $\cf: L^2 (G) \rightarrow L^2_\star(\fg^*)$ between the representations, which can be expressed as an integral transform. Namely,
\ba
	\tilde\psi(X):= \cf(\psi)(X) = \int_G \d g\, E(g,X)\, \psi(g) \in L^2_\star(\fg^*)\,, \nonumber
\ea
where $\psi \in L^2(G)$, and we denote by $E(g,X)$ the integral kernel of the transform. Then, the goal is to identify the defining equations for the kernel $E(g,X)$ using the fact that the intertwined function spaces define a representation of the same quantum algebra, and applying the action of $\fA$ in the different representations. If a solution exists, we will have thus shown that the representations are related through the corresponding integral transform. Once more, its actual existence has to be verified once an explicit choice of quantization map and $\star$-product has been made.

\

The intertwining property of $\cf$ can be expressed generally as $\cf \circ \pi_G(\hat{T}) = \pi_{\fg^*}(\hat{T}) \circ \cf$, where $\hat{T}\in\fA$. For the $\hat{X}_i$ operators we have
\ba
	\cf(\pi_G(\hat{X}_i)\psi)(X) &= \int_G \d g\, E(g,X)\, (i\cl_i\psi)(g) \nn
	&= \int_G \d g\, (-i\cl_i E)(g,X)\, \psi(g) \,, \nonumber
\ea
where for the last equality we used integration by parts and $\psi \in L^2(G)$.
On the other hand,
\ba
	(\pi_{\fg^*}(\hat{X}_i)\cf(\psi))(X) = \int_G \d g\, (X_i \star E(g,X))\, \psi(g) \,, \nonumber
\ea
and accordingly, for all $\psi\in L^2(G)$ we must require the kernel $E(g,X)$ to satisfy the differential equation
\ba\label{eq:LE}
	-i\cl_i E(g,X) = X_i \star E(g,X) \,.
\ea
Integrating this action by right-invariant Lie derivatives, we obtain
\ba\label{eq:EgX}
	E(hg,X) = e^{k(h)\cdot\vec{\cl}}E(g,X) = e_\star^{ik(h)\cdot X} \star E(g,X) \,,
\ea
where again $k(h) = -i\ln(h) \in \fg$, and we introduced the $\star$-exponential notation
\ba
	e_\star^{f(X)} = \sum_{n=0}^{\infty} \frac{1}{n!} \underbrace{f \star \cdots \star f}_{n\ \textrm{times}}(X) \,. \nonumber
\ea

Of course, such an integration of a differential equation is subject to the possible non-trivial global properties of $G$. First of all, the assumption that $G$ is exponential guarantees that any group element $h$ can be integrated to as in (\ref{eq:EgX}). However, since $E(g,X)$ is to be considered only under integration, weak exponentiality of $G$ is a sufficient condition for our purposes. On the other hand, if $G$ has compact subgroups, the logarithm map is multivalued, and therefore the result of the integration is not unique. In particular, we may choose $k(h) = -i\ln(h)\in\fg$ from any branch of the logarithm, each one supplying a solution of the differential equation (\ref{eq:LE}).

Consider then the intertwining of the operators $\hat{\zeta}^i$. We have
\ba
	\cf(\pi_G(\hat{\zeta}^i)\psi)(X) = \int_G \d g\, E(g,X)\,\zeta^i(g)\,\psi(g) \, \nonumber
\ea
and, on the other hand,
\ba
	(\pi_{\fg^*}(\hat{\zeta}^i)\cf(\psi))(X) = \int_G \d g\, (-i\del^iE)(g,X)\, \psi(g) \, \nonumber
\ea
for all $\psi\in L^2(G)$. We must therefore require
\ba
	(-i\del^iE)(g,X) = \zeta^i(g) E(g,X) \,,
	\label{eq:DE2}
\ea
which through integration yields
\ba\label{eq:EgXpY}
	E(g,X+Y) = e^{Y\cdot\vec{\del}}E(g,X) = e^{i\zeta(g)\cdot Y}E(g,X) \,.
\ea
Since $\fg\cong\Rl^d$, there are no global issues with this integration. Here the multivaluedness comes in through the possible multivaluedness of the coordinates $\zeta:G\rightarrow\fg$.

From (\ref{eq:EgXpY}) we have, in particular, that $E(e,X) = E(e,0)=:c$ is constant in the principal branch, since $\zeta_k(0)=0$. We will set $c\equiv 1$. Combining this with (\ref{eq:EgX}), we find
\ba
E(g,X) = e_\star^{ik(g)\cdot X} \,,
\label{eq:Estar}
\ea
where again $k(g)=-i\ln(g)$ may a priori be taken from any branch of the logarithm. 
Thus, given a suitable deformation quantization $\star$-product, this formula gives the general expression for the integral kernel $E(g,X)$. 

However, we also find from (\ref{eq:EgXpY}) another form
\ba
E(g,X) = \eta(g) e^{i\zeta(g)\cdot X}
\label{eq:Eetaexp}
\ea
for the kernel. The prefactor $\eta(g):=E(g,0)$ may be non-trivial depending on the $\star$-product or, equivalently, the quantization map $\cq$ chosen, as we will see in Section \ref{sec:app}.\\

Let us note that the expressions (\ref{eq:Estar}) and (\ref{eq:Eetaexp}) are, in fact, solutions to two distinct differential equations (\ref{eq:LE}) and (\ref{eq:DE2}), respectively, and for consistency we must require them to define the same function. Of course, for a given $\star$-product, determining coordinates for which this equality is satisfied might be a difficult task and, in general, there is no guarantee that such coordinates exist. It is a consistency requirement for the non-commutative Fourier transform to arise as an intertwiner between the group representation and the algebra representation. In fact, as we will see in \ref{ssec:G&alg}, the algebra representation is only guaranteed to exist under the conditions that such coordinates can be found, tying together the \emph{existence} of the non-commutative Fourier transform as an intertwiner with that of the algebra representation, and \emph{vice versa}.

Accordingly, for a given $\star$-product, the last two equations give the explicit form of the corresponding plane waves. They signify two important things. First, the non-commutative plane waves take generically the form of $\star$-exponentials with respect to the $\star$-product (following from the quantization map $\cq$) in terms of the canonical coordinates $k(g)$ on the group. That is, they are obtained by the inverse quantization map $\cq^{-1}$ applied to the operators $e^{i k(g) \cdot \hat{X}} \in \fA_{\fg^*}$. Second, under the above consistency requirement that (\ref{eq:Estar}) defines the same function as (\ref{eq:Eetaexp}), there exists a choice of coordinates $\zeta^i(g)$, in which the same $\star$-exponentials take the form of classical exponentials times a multiplicative factor $\eta(g)$. Also, the preferred coordinates on the group and the measure factor that appear in this last expression thus follow uniquely from the choice of quantization map together with the $\star$-product.\\

Let us now note a very important point. From (\ref{eq:Estar}) we have that $\cq(E(g,X)) = e^{ik(g)\cdot\hat{X}} \in \fA_{\fg^*} \cong \overline{U(\fg)}$, where $k(g)=-i\ln(g)\in\fg$, and the quantization map is applied only to the coordinates $X_i$ on $\fg^*$. Elements of this form in $\fA_{\fg^*}$ constitute a group: Since $\hat{X}_i$ obey the Lie algebra commutation relations, we have
\ba
	e^{ik\cdot\hat{X}}e^{ik'\cdot\hat{X}} = e^{i\cb(k,k')\cdot\hat{X}} \,,\nonumber
\ea
where $\cb(k,k')$ is obtained through Baker-Campbell-Hausdorff formula, and $k,k'\in\fg$. Let us denote this group by $\ce:=\{e^{ik\cdot\hat{X}}:k\in\fg\} \subset \fA_{\fg^*}$. However, because of the possible multivaluedness of the logarithm, there is in general no one-to-one relation between the elements of $\ce$ and the group $G$. The Lie algebra element $k(g)$ may lie in any branch of the multivalued logarithm, and the Baker-Campbell-Hausdorff formula applied to Lie algebra elements in one branch need not lie in the same branch. As already noted before, there is in particular a set of elements $\ci:=\{e^{ik\cdot\hat{X}} \in \fA_{\fg^*}: e^{k\cdot\vec{\cl}} = 1\} \subset \ce$, which correspond to translations around compact subgroups of $G$ in the group representation. In fact, $\ci$ is a normal subgroup of $\ce$, so we may consider the quotient group $\ce/\ci$, which is then isomorphic to $G$ itself (assuming again that $G$ is exponential), because the different branches of the logarithm are thus identified. Therefore, it would be natural to define the non-commutative plane waves as the equivalence classes of elements $E_g(X) := \{e_\star^{ik\cdot X} \in C^\infty(\fg\times\fg^*): k=-i\ln(g)\}$, which is the straightforward translation of the above quotient group to $\star$-exponentials. $E_g(X)$ then constitute a representation of $G$ under $\star$-multiplication. However, for practical purposes, it is more convenient and transparent simply to introduce a new product `$\star_p$' for non-commutative plane waves, which is the deformation quantization $\star$-product amended by a projection onto the principal branch of the logarithm. In a sense, this new product sees the global structure of $G$, whereas the deformation quantization $\star$-product is a purely local construct arising from the Lie algebra alone. (For the action of the generators of $\fA$ in the different representations above we considered only infinitesimal translations, which are unaffected by global properties of $G$.) Then, we define
\ba
E_g(X) := e_\star^{ik(g)\cdot X}\,, 
\ea
\emph{where $k(g)=-i\ln(g)\in\fg$ is taken in the principal branch; and constitute a representation of $G$ with respect to the $\star_p$-product.} For weakly exponential Lie groups a representation is obtained in a weak sense.\\

\

With the remarks from above on the coordinates $\zeta^i(g)$, let us then list some important properties of the non-commutative plane wave $E_g(X)$, as they follow from our construction, which we will use in the following:
\ba
E_g(X) &= e_\star^{ik(g)\cdot X}\,=\, \eta(g)e^{i\zeta(g)\cdot X}\,, \\
E_e(X) &= 1\,, \\
\cq(E_g(X)) &= e^{ik(g)\cdot\hat{X}}\in\fA_{\fg^*}\,, \\ 
E_{g^{-1}}(X) &= \overline{E_g(X)} = E_g(-X)\,, \\
\label{eq:planewavescomposition}
E_{gh}(X) &= E_g(X) \star_p E_h(X)\,.
\ea
In addition, using
\bas
	E_{g}(X) \equiv \eta(g) e^{i\zeta(g)\cdot X},\ \eta(e)=E_e(0)\equiv 1\,,
\eas
and the properties of the $\zeta$-coordinates, namely, $\zeta(e)=0$ and $\cl_i\zeta^j(e)=\delta_i^j$, we have
\ba
\label{eq:EGdelta}
\int_{\fg^*}\frac{\d^dX}{(2\pi)^d}\, E_{g}(X) = \delta^d(\zeta(g)) = \delta(g) \,,
\ea
where the right-hand side is the Dirac delta distribution with respect to the right-invariant Haar measure on $G$.\\

We have thus found an integral transform $\cf$ intertwining the representations $\pi_G$ and $\pi_{\fg^*}$:
\ba
	\tilde{\psi}(X):= \cf(\psi)(X) = \int_G \d g\, e_\star^{ik(g)\cdot X}\, \psi(g) \,,
\ea
where $k(g)=-i\ln(g)$ is taken in the principal branch. The $\star_p$-product of non-commutative plane waves is extended by linearity to the image of $\cf$.\\

\subsection{Properties of the non-commutative Fourier transform}
\label{ssec:propF}

Let us now consider some properties of the transform $\cf$ and the non-commutative function space $L_\star^2(\fg^*)$:
\begin{itemize}
\item Group multiplication from the right is dually represented on $\cf(\psi)(X)$ as $\star_p$-multiplication by $E_{g^{-1}}(X)$, i.e.,
\bas
	\cf(R_g\psi)(X) &= \int_G\d h\, E_h(X)\, \psi(gh) \nn
	&= \int_G\d h\, E_{g^{-1}h}(X)\,\psi(h) \nn
	&= E_{g^{-1}}(X) \star_p \int_G\d h\, E_{h}(X)\, \psi(h) \nn
	&= E_{g^{-1}}(X) \star_p \cf(\psi)(X)
\eas
using the right-invariance of the Haar measure.

\item Consider the $L_{\star}^2(\fg^*)$ inner product of two functions obtained through the transform
\bas
	\langle \tilde{\psi},\tilde{\psi}'\rangle_{\fg^*} &:= \int_{\fg^*}\frac{\d^dX}{(2\pi)^d}\, \overline{\tilde{\psi}(X)} \star_p \tilde{\psi}'(X) \nn
	&= \int_{\fg^*}\frac{\d^dX}{(2\pi)^d} \left[\int_G \d g\, E_{g^{-1}}(X)\, \overline{\psi(g)} \right] \star_p \left[\int_G \d h\, E_h(X)\, \psi'(h)\right] \nn
	&= \int_G \d g\, \int_G \d h\, \overline{\psi(g)}\, \psi'(h) \left[\int_{\fg^*}\frac{\d^dX}{(2\pi)^d} E_{g^{-1}h}(X) \right] \,.
\eas
Using (\ref{eq:EGdelta}), we find
\bas
	\langle \tilde{\psi},\tilde{\psi}'\rangle_{\fg^*} &\equiv \int_{\fg^*}\frac{\d^dX}{(2\pi)^d}\, \overline{\tilde{\psi}(X)} \star_p \tilde{\psi}'(X) = \int_G \d g\, \overline{\psi(g)}\, \psi'(g) \equiv \langle \psi,\psi'\rangle_G \,,
\eas
so $\cf$ is, in fact, an isometry from $L^2(G)$ to $L_{\star}^2(\fg^*)$. Therefore, we may identify $L_{\star}^2(\fg^*) = \cf(L^2(G))$.

\item Consider the transformation $\cf^*: L_\star^2(\fg^*) \rightarrow L^2(G)$ given by
\ba\label{eq:Finverse}
	\cf^*(\tilde\psi)(g) := \int_{\fg^*}\frac{\d^dX}{(2\pi)^d}\, \overline{E_g(X)} \star_p \tilde\psi(X) \,.
\ea
We have
\bas
	(\cf^* \circ \cf)(\psi)(g) &= \int_{\fg^*}\frac{\d^dX}{(2\pi)^d}\, \overline{E_g(X)} \star_p \int_G \d h\, E_h(X)\,\psi(h) \nn
	&= \int_G \d h\, \left[\int_{\fg^*}\frac{\d^dX}{(2\pi)^d}\, E_{g^{-1}h}(X) \right]\psi(h) \nn
	&= \int_G \d h\, \delta(g^{-1}h)\, \psi(h) = \psi(g) \,.
\eas
That is, $\cf^* \circ \cf = \text{id}_{L^2(G)}$.

\item For $\cf \circ \cf^*$ we find
\bas
	(\cf \circ \cf^*)(\tilde\psi)(X) &= \int_G\d g\, E_g(X) \int_{\fg^*}\frac{\d^dY}{(2\pi)^d}\, \overline{E_g(Y)} \star_p \tilde\psi(Y) \nn
	&= \int_{\fg^*}\frac{\d^dY}{(2\pi)^d}\, \left[\int_G\d g\, E_g(X) E_g(-Y) \right] \star_p \tilde\psi(Y) \,,
\eas
which shows that the (generalized) function
\ba
	\delta_\star(X,Y) := \int_G\d g\, E_g(X) E_g(-Y) \in (L_\star^2(\fg^*))^*
\ea
acts as the integration kernel of the projection operator $\cf \circ \cf^*$ onto $L_\star^2(\fg^*)$ (with respect to the $\star_p$-product), and accordingly corresponds to the Dirac delta in $L_\star^2(\fg^*)$. 

\item It is easy to check that the kernel of $\cf \circ \cf^*$, $\ker(\cf \circ \cf^*) = \{ \tilde\psi \in L^2_\star(\fg^*): (\cf \circ \cf^*)(\tilde\psi)=0\}$, contains all functions of the form $(e_\star^{ik(e)\cdot X} - e_\star^{ik'(e)\cdot X})\, \star\, \tilde\psi(X),\ \tilde\psi \in L^2_\star(\fg^*)$, where $k(e),k'(e)\in\fg$ are any two values of $-i\ln(e)$, and therefore $\cf \circ \cf^*$ implements the aforementioned $\ce/\ci$-equivalence classes in $L_\star^2(\fg^*)$.

\item We have an expression (or two) for the $\star_p$-product under integration in terms of a pseudo-differential operator $\sigma$, namely,
\ba
	\int_{\fg^*} \d^d X\ \overline{\tilde{\psi}(X)} \star_p \tilde{\psi'}(X) &= \int_{\fg^*} \d^d X\  \left(\sigma(i\vec{\prt})\, \overline{\tilde{\psi}(X)}\right) \tilde{\psi'}(X) \nn 
	&= \int_{\fg^*} \d^d X\ \overline{\tilde{\psi}(X)} \left(\sigma(-i\vec{\prt})\, \tilde{\psi'}(X)\right) \label{eq:starint}
\ea
$\forall\, \tilde{\psi},\tilde{\psi'} \in L_\star^2(\fg^*)$, where $\sigma(\zeta) := \big(\omega(\zeta)|\eta(\zeta)|^2\big)^{-1}$ for $\zeta\in\fg$, $\d g \equiv \omega(\zeta(g))\,\d \zeta(g)$ for the right-invariant Haar measure, and $\eta(\zeta(g)) \equiv E(g,0)$. For the proof of this identity refer to the Appendix \ref{app:starint}.

\item Due to (\ref{eq:starint}), we may write the inverse transform $\cf^{-1} \equiv \cf^*: L_\star^2(\fg^*) \rightarrow L^2(G)$ from (\ref{eq:Finverse}) explicitly without a star-product as
\ba\label{eq:Finvsigma}
	\cf^{-1}(\tilde\psi)(g) = \sigma(g) \int_{\fg^*}\frac{\d^dX}{(2\pi)^d}\, \overline{E_g(X)}\, \tilde\psi(X) \,,
\ea
where $\sigma(g) := \big(\omega(\zeta(g))|\eta(g)|^2\big)^{-1}$.

\item Finally, due to $E_g\star_p E_h=E_{gh}$, the $\star_p$-product is dual to the convolution product on $G$ under the non-commutative Fourier transform, i.e.,
\ba
\tilde{\psi}\star_p \tilde{\psi}' = \widetilde{\psi * \psi'}\,,
\ea
where the convolution product is defined on the group as usual
\bas
\psi * \psi' (g) =\int_G \d h\, \psi(gh^{-1})\psi'(h)\,.
\eas
\end{itemize}

\

Let us emphasize again the difference to standard harmonic analysis on locally compact groups: In that case the Peter-Weyl theorem would take us through the expansion of functions on $G$ in terms of unitary irreducible representations, and the Fourier transform would give us a unitary map from square-integrable functions $L^2(G)$ on the group $G$ to square-integrable functions $L^2(\widehat{G})$ on the Pontryagin dual $\widehat{G}$:
\bas
\hat{\psi}_\lambda :=\int_G \d g\, \psi(g)\, \rho_\lambda(g^{-1})\,,\\
\psi(g) := \sum_{\lambda\in \widehat{G}}d_\lambda \text{Tr}[\hat{\psi}_\lambda\, \rho_\lambda(g)]\,,
\eas
where $\rho_\lambda(g)$ is a unitary irreducible representation of $G$ on a vector space of dimension $d_\lambda$. Note that in the special case of Euclidean space the Pontryagin dual $\widehat{G}$ happens to coincide with the momentum space $\fg^*$, and therefore the non-commutative Fourier transform and the Fourier transform coming from the Peter-Weyl theorem coincide, as discussed in section \ref{sec:motivation}. Nevertheless, let us also note, that in the context of locally compact Lie groups we will have both transforms at our disposal.

\subsection{Compatible coordinates and existence of algebra representation}
\label{ssec:G&alg}
As an aftermath of the derived form and properties of the non-commutative plane wave and the corresponding interwiner of the representations $\pi_G$ and $\pi_{\fg^*}$ --- the non-commutative Fourier transform $\cf$ ---, let us inquire a bit further on the existence of the algebra representation $\pi_{\fg^*}$ for a specific choice of coordinates on the group $G$. Recall the property (\ref{eq:leibnizg}) encoding the compatibility between a $\star$-product (or, equivalently, a quantization map) and a choice of coordinates on the group, which follows from the coproduct structure of the quantum algebra of observables to be represented, and is needed for the existence of an algebra representation of the same. This was also represented as the commutative diagram (\ref{eq:leibnizCD}). Given the coordinates $\zeta:G\rightarrow \fg \cong \Rl^d$ on $G$ arising from the star-exponential of the non-commutative plane wave as $E_g(X) = e_\star^{i k(g)\cdot X} = \eta(g) e^{i\zeta(g)\cdot X}$, \emph{determined by a suitable $\star$-product leading to such a form}, they compose as
\bas
	\zeta^i(gh) &= \sum_{n=1}^{\infty} \sum_{\substack{k,l\in\Nl\\k+l=n}} C^i_{p_1\cdots p_k q_1\cdots q_l} \zeta^{p_1}(g) \cdots \zeta^{p_k}(g) \zeta^{q_1}(h) \cdots \zeta^{q_l}(h) =: \cc(\zeta(g),\zeta(h))^i \,,
\eas
where $C^i_{p_1\cdots p_k q_1\cdots q_l}\in \Rl$ are constant coefficients. This gives rise to the following coproduct, as in (\ref{eq:zetacoprod}),
\bas
\Delta_G(\hat{\zeta}^i) &=  \cc(\hat{\zeta}_{(1)},\hat{\zeta}_{(2)})^i \equiv \sum_{n=1}^{\infty} \sum_{\substack{k,l\in\Nl\\k+l=n}} C^i_{p_1\cdots p_k q_1\cdots q_l} \hat{\zeta}^{p_1} \cdots \hat{\zeta}^{p_k} \otimes \hat{\zeta}^{q_1} \cdots \hat{\zeta}^{q_l} \,,
\eas
where the lower indices $(1),(2)$ refer to the first and the second factor on the tensor product, on which the coproduct operates. In the algebra representation this yields
\bas
(\pi_{\fg^*}\otimes \pi_{\fg^*})(\Delta_G(\hat{\zeta}^i)) = \cc(-i\vec{\prt}_{(1)},-i\vec{\prt}_{(2)})^i \,.
\eas
Now, for a given $\star$-product, we want to check the commutativity of the diagram (\ref{eq:leibnizCD}), i.e., that Equation (\ref{eq:leibnizg})
\bas
m_{\fg^*} \circ (\pi_{\fg^*}\otimes\pi_{\fg^*})(\Delta_{G}(\hat{\zeta}^i)) = \pi_{\fg^*}(\hat{\zeta}^i) \circ m_{\fg^*}
\eas
is satisfied. It will be enough to do the calculation at the level of the exponentials, once a Fourier transform is established, since any function can then be written in terms of them. This can be done by explicit calculation for exponentials. We want to show that
\ba\label{eq:conditionPW}
	& m_{\fg^*} \circ (\pi_{\fg^*}\otimes\pi_{\fg^*})(\Delta_{G}(\hat{\zeta}^i))(E_{g_1}(X) \otimes E_{g_2}(X)) \nn
	&= \pi_{\fg^*}(\hat{\zeta}^i) \circ m_{\fg^*}(E_{g_1}(X)\otimes E_{g_2}(X)) \,.
\ea
The left-hand side of (\ref{eq:conditionPW}) reads explicitly
\bas
&m_{\fg^*} \circ (\pi_{\fg^*}\otimes\pi_{\fg^*})(\Delta_{G}(\hat{\zeta}^i))(E_{g_1}(X) \otimes E_{g_2}(X)) \nn
&= m_{\fg^*} (\cc(-i\vec{\prt}_{(1)},-i\vec{\prt}_{(2)})^i\, E_{g_1}(X) \otimes E_{g_2}(X)) \nn
&= m_{\fg^*} (\cc(\zeta(g_1),\zeta(g_2))^i\, E_{g_1}(X) \otimes E_{g_2}(X)) \nn
&=\zeta^i(g_1g_2) E_{g_1g_2}(X)\,,
\eas
where we used $-i\prt^i E_g(X) = \zeta^i(g) E_g(X)$. Similarly, the right-hand side of (\ref{eq:conditionPW}) reads:
\bas
&\pi_{\fg^*}(\hat{\zeta}^i) \circ m_{\fg^*}(E_{g_1}(X)\otimes E_{g_2}(X)) = -i\prt^i E_{g_1g_2}(X) = \zeta^i(g_1g_2)E_{g_1g_2}(X)\,,
\eas
thus proving the equality. Accordingly, we see that \emph{when the $\star$-product is verified to lead to a non-commutative plane wave of the form} $E_g(X)=\eta(g)e^{i\zeta(g)\cdot X}$, as it happens in all the examples we will consider below, then it is guaranteed that the $\zeta$-coordinates in the exponential, along with their coproduct, are compatible with the $\star$-product in the sense of the commutative diagram (\ref{eq:leibnizCD}).\footnote{It is an interesting question, which we will not address here, whether the non-commutative plane wave must be of the above form in order for a compatible coordinate system to exist, and furthermore, how to characterize the class of star-products, for which such coordinates can be found.}

\section{Explicit examples}
\label{sec:app}
We have seen that the $\star$-product used in defining the algebra representation follows from the choice of quantization map, by the formula (\ref{eq:definitionstar}). Further, the key ingredient needed for the definition of the non-commutative Fourier transform is the non-commutative plane wave. This can be computed explicitly as soon as a quantization map (ordering prescription) for the algebra coordinate operators (equivalently, a $\star$-product) is chosen, such that it allows for a compatible set of coordinates as encoded in the diagram (\ref{eq:leibnizCD}).

We will now provide a few explicit examples of our construction. We start from the rather trivial, but still interesting, abelian $\U(1)$ case (also considered in \cite{carlosbiancadaniele}), and then move on to the non-abelian but still compact $\SU(2)$ case. In the latter we consider three quantization maps: the symmetric map (corresponding to the Weyl ordering), the Duflo map, and the so-called Freidel-Livine-Majid map. The corresponding $\star$-products and non-commutative plane waves are computed and shown to be of the form required for the existence of the algebra representation, in particular, $E_g(X)=\eta(g)e^{i\zeta(g)\cdot X}$ as proved above in Subsection \ref{ssec:G&alg}. Finally, the non-commutative Fourier transforms along with their inverses are presented.\\

Before considering each of the following examples let us show how, in practice, one determines the non-commutative plane waves. Recall that the plane wave is given by
\bas
E_g(X) &=e_\star^{ik(g)\cdot X}=\sum_{n=0}^\infty\frac{i^n}{n!}k(g)^{i_1}\cdots k(g)^{i_n}X_{i_1}\star\cdots\star X_{i_n}\nn
&=\sum_{n=0}^\infty\frac{i^n}{n!}k(g)^{i_1}\cdots k(g)^{i_n}\cq^{-1}(\hat{X}_{i_1}\cdots \hat{X}_{i_n})\nn
& = \cq^{-1}(e^{ik(g)\cdot \hat{X}})\,.
\eas
Therefore, in order to obtain the explicit form of the plane waves, one can either compute the inverse quantization map for all the monomials, or one can guess which function upon quantization gives $e^{ik(g)\cdot \hat{X}}$, that is, the function $f(X)$ such that $\cq(f(X))=e^{ik(g)\cdot \hat{X}}$. (Notice that $\cq^{-1}(\hat{X}_{i_1}\cdots \hat{X}_{i_n})\neq \cq^{-1}(\widehat{X_{i_1}\cdots X_{i_n}})=X_{i_1}\cdots X_{i_n}$.) As we will see, for the examples we will present, this latter route turns out to be the most straightforward.
Besides, once $E_g(X)$ is known, by using the property (\ref{eq:planewavescomposition}), $E_{g_1}\star_p E_{g_2}=E_{g_1g_2}$, one can determine the $\star$-product on monomials as
\ba
\label{eq:monomialsstar}
X_{i_1}\star\cdots \star X_{i_n} &=(-i)^n\,\left.\frac{\del^n}{\del k(g_1)^{i_1}\cdots \del k(g_n)^{i_n}}\right|_{g_1,\ldots,g_n=e}\,\cq^{-1}(e^{ik(g_1\cdots g_n)\cdot \hat{X}}) \nn
&\equiv \left. (-i)^n \cl_{i_1} \cdots \cl_{i_n} E_g(X) \right|_{g=e}\, , 
\ea
thus reverting, in some sense, the natural logic of the construction. (Notice that the projection in $\star_p$ is of no consequence in this formula, since the derivatives are taken in the neighborhood of the identity.)

\subsection{Commutative: $\U(1)$}
$\U(1)$ is given by the set of complex numbers $z\in\Cl$ with modulus one $|z|=1$. Accordingly, we can set $z=e^{i\theta}$. The canonical coordinates $k(g) = -i\ln(g) \equiv \theta$ are restricted to the  principal branch of the logarithm as $\theta\in]-\pi,\pi]$. The dual of the Lie algebra $\fu(1)^*$ is simply given by the real numbers $X\in\Rl$.

In this abelian case, and in particular for $\fu(1)$, which has just one generator, no ordering ambiguity arises, so that there is no difference between quantization maps in this respect. However, first of all, the group is compact, and this topological feature already makes things a little more interesting. Second, we have seen how the quantization map also affects the choice of coordinates appearing in the plane waves. It is then worth to consider this simple case in some detail.

For the symmetrization map $\cs$, Equation (\ref{eq:symmetrizationmap}), (and also for the Duflo map $\cd$ which we will consider below, as they coincide for abelian groups) we indeed have $\cs(X^n)=\hat{X}^n$ and, therefore,
\bas
\cs(e^{i\theta X})=e^{i\theta \hat{X}}\,,
\eas
that is, as expected, the plane waves are given by $e^{i\theta X}$, for $\theta\in]-\pi,\pi]$, $X\in\Rl$, and the corresponding $\star$-product on monomials is simply the pointwise product
\bas
 \underbrace{X\star \cdots \star X}_{n\ \mathrm{times}} = X^n\,.
\eas
Nevertheless, the product $e^{i\theta X} \cdot_p e^{i\theta' X} = e^{i(\theta+\theta' (\textrm{mod}\ 2\pi)) X}$ of plane waves is still non-trivial due to the compactness of the group, which has to be taken into account by explicit projection, as we explained above in the general case.

Furthermore, from (\ref{eq:starint}) we have that
\bas
	\int \d X\ f(X) \cdot_p f'(X) = \int \d X\ f(X) f'(X) \,,
\eas
since in this case $\d g = \d\theta \Rightarrow \omega(\theta)=1$ and $E_g(X) = e^{i\theta X} \Rightarrow \eta(\theta)=1$, so $\sigma=1$.

The non-commutative Fourier transform is thus given by
\ba
\label{eq:u1fourier}
\tilde{\psi}(X)=\int_{-\pi}^{\pi}\d\theta \,e^{i\theta X}\,\psi(e^{i\theta})\,,
\ea
while its inverse is
\ba
\label{eq:u1inv}
\psi(e^{i\theta})=\int_\Rl \frac{\d X}{2\pi}\, e^{-i\theta X}\,\tilde{\psi}(X)\,.
\ea

Let us now point out one consequence of the existence of normal subgroups corresponding to the identity element in this simple case. The periodicity of the group is taken care of by the projection in the product $\cdot_p$, which translates it into the equivalence class of functions on the Lie algebra $\tilde{\psi}(X) = e^{i2\pi n X}\cdot_p\tilde{\psi}(X)$, $n\in\Zl$. This is the counterpart, in our setting, of the usual Fourier transform on the circle, where the restriction $X\in\Zl$ is imposed, and the inverse transform is given by a sum over the integers.

In fact, it was proved in \cite{carlosbiancadaniele} that this $\U(1)$ non-commutative Fourier transform defined for the full $\Rl$ can, in fact, be determined by its values on the integers; thus, even though the $\U(1)$ non-commutative Fourier transform is defined distinctively from the usual Fourier transform on the circle, they were shown to coincide due to this form of sampling.\\

We have thus seen that the symmetric (and Duflo) map leads to plane waves equivalent to the usual ones. Still, we have also seen within the general formalism that the choice of quantization maps affects non-trivially also the coordinates appearing in the plane waves. Vice versa, by choosing non-linear coordinates on the group, one can end up with non-trivial star-products, despite the abelianess of the group. Let us say we have $\cq$ such that
\bas
\cq^{-1}(e^{i\theta \hat{X}})=e^{2i\sin\frac{\theta}{2} X}\,.
\eas
$\zeta(\theta) = 2\sin\frac{\theta}{2}$ can be seen as new coordinates on the group valid for $\theta\in]-\pi,\pi]$. According to (\ref{eq:monomialsstar}), we get for the $\star$-product on monomials already a diverting result at third order
\bas
X\star X &= X^2\,,\\
X\star X\star X &= X^3+\frac{1}{4}X\,,\\
\vdots\nonumber
\eas
Of course, we still have $f \star_p f' - f' \star_p f = 0$ for all $f,f'\in C^\infty(\Rl)$, so that the (trivial) Lie algebra relations are well-represented, and $\star$ is a genuine deformation quantization star-product. Therefore, as remarked before, we see that quantization map, choice of coordinates, and star-product are related in a highly non-trivial way.

We may give an expression for the corresponding $\star_p$-product under integral, from (\ref{eq:starint}), as a (non-trivial) pseudo-diffential operator
\ba\label{eq:Weirdostarint}
	\int \d X\ f(X) \star_p f'(X) = \int \d X\ f(X)\, \sqrt{\textstyle 1+\textstyle\frac{1}{4}\left(\frac{\d}{\d X}\right)^2}\, f'(X) \,,
\ea
(where $\frac{\d}{\d X}$ may act either left or right) as we now have, in contrast to the previous parametrization, a non-trivial relation between the Haar measure $\d\theta$ and the Lebesgue measure $\d\zeta$, namely, $\d\theta=(\sqrt{1-\zeta^2/4})^{-1}\d\zeta$, so $\sigma(\zeta) = \sqrt{1-\zeta^2/4}$.

The non-commutative Fourier transform is thus given by
\ba
\tilde{\psi}(X)=\int_{-\pi}^{\pi}\d\theta\ e^{2i\sin\frac{\theta}{2} X}\,\psi(e^{i\theta})\,,
\ea
while its inverse is, from (\ref{eq:Finvsigma}), 
\ba
\psi(e^{i\theta})=\cos({\textstyle\frac{\theta}{2}})\int_\Rl \frac{\d X}{2\pi}\ e^{-2i\sin\frac{\theta}{2} X} \, \tilde{\psi}(X)\,.
\ea

\subsection{Non-commutative compact: $\SU(2)$}
We now consider a simple but very important non-abelian example, $\SU(2)$, which is particularly relevant also for quantum gravity applications. The Lie algebra $\su(2)$ has a basis given (in the defining representation) by a set of two-by-two traceless hermitian matrices $\{\sigma_j\}_{j=1,2,3}$, which read
\bas
\sigma_1= \left( \begin{array}{cc}
0 & 1 \\
1 & 0  \end{array} \right)\,,\q
\sigma_2= \left( \begin{array}{cc}
0 & -i \\
i & 0  \end{array} \right)\,,\q
\sigma_3= \left( \begin{array}{cc}
1 & 0 \\
0 & -1  \end{array} \right)\,,
\eas
and satisfy $\sigma_i\sigma_j=\delta_{ij}+i\eps_{ijk}\sigma_k$. Thus, a generic element $k\in\su(2)$ can be written as $k=k^j\sigma_j$, $k^j\in\Rl$, while for any group element $g\in\SU(2)$ we may write $g=e^{ik^j\sigma_j}$. Thus, $\SU(2)$ is an exponential Lie group. Another convenient parametrization of $\SU(2)$ can be written as
\ba
g=p^0\1+ip^i\sigma_i\,,\q (p^0)^2+p^ip_i=1\,,\q p^i\in\Rl\,.
\ea 
Here, the $p^i$'s are constrained by the $\Rl^3$ vector norm $|\vec{p}|^2\leq1$. Thus, this last parametrization naturally identifies $\SU(2)$ with the 3-sphere $S^3$. $p^0\geq 0$ and $p^0\leq 0$ correspond to the upper and lower hemispheres of $S^3$, respectively, in turn corresponding to two copies of $\SO(3)$. Parametrization of the group elements in terms of $\vec{p}\in\Rl^3$ is one-to-one only on either of the two hemispheres, whereas the canonical coordinates $\vec{k}$ parametrize the whole group except for $-\1\in \SU(2)$.

The relation between these two parametrizations is mediated by the following change of coordinates
\ba
\label{eq:pparametrization}
\vec{p}=\frac{\sin|\vec{k}|}{|\vec{k}|}\vec{k}\,,\q p_0=\cos|\vec{k}|\,,\qquad k^i\in\Rl\,,
\ea
where $|\vec{k}|\in[0,\frac{\pi}{2}[$, or $|\vec{k}|\in[\frac{\pi}{2},\pi[$ according to $p^0\geq 0$, $p^0\leq 0$ respectively, and $g\in\SU(2)$ assumes the form
\bas
g=\cos|\vec{k}|\1+i\frac{\sin|\vec{k}|}{|\vec{k}|}\vec{k}\cdot\vec{\sigma}=e^{i \vec{k}\cdot \vec{\sigma}}\,.
\eas

We call the coordinates introduced the $\vec{k}$-parametrization and the $\vec{p}$-parametrization, respectively. The Haar measure on the group takes then the form
\ba
\d g &= \d^3 \vec{k}\left(\frac{\sin|\vec{k}|}{|\vec{k}|}\right)^2\,,\q \vec{k}\in\Rl^3\,, |\vec{k}|\in [0,\pi[\,, \label{eq:kHaar}\\
\d g &= \frac{\d^3 \vec{p}}{\sqrt{1-|\vec{p}|^2}}\,,\q \vec{p}\in \Rl^3\,, |\vec{p}|^2<1\,, \label{eq:pHaar}
\ea
where the latter is again applicable only for one of the two hemispheres.\\

We now consider three choices of quantization maps, and derive the corresponding $\star$-product, algebra representation and non-commutative plane waves.

\subsubsection{Symmetrization map}
Given a set of $\su(2)$ coordinates $X_{i_1},\ldots,X_{i_n}$, the symmetrization map $\cs$ takes the symmetric ordering of the corresponding coordinate operators $\hat{X}_{i_1},\ldots,\hat{X}_{i_n}$,
\bas
\mathcal{S}(X_{i_1}\cdots X_{i_n}) = \frac{1}{n!}\sum_{\sigma\in S_n} \hat{X}_{i_{\sigma_1}}\cdots \hat{X}_{i_{\sigma_n}}\,,
\eas
where $S_n$ is the symmetric group of order $n$. Thus, for instance, for an exponential of the form $e^{i\vec{k}\cdot {X}}$, we have
\bas
\mathcal{S}(e^{i\vec{k}\cdot {X}}) &=\1+ik^{i}\cs({X}_{i})+\frac{i^2k^ik^j}{2!}\cs({X_iX_j})+\frac{i^3k^ik^jk^k}{3!}\cs({X_iX_jX_k})+\ldots\nn
&=\1+ik^i \hat{X}_i+\frac{i^2k^ik^j}{2!}\frac{1}{2!}(\hat{X}_i\hat{X}_j+\hat{X}_j\hat{X}_i)+\ldots\nn
&=\1+ik^i \hat{X}_i+\frac{i^2k^ik^j}{2!}\hat{X}_i\hat{X}_j+\ldots\nn
&\equiv e^{i\vec{k}\cdot \hat{{X}}}\,,
\eas
which tells that the function $e^{i\vec{k}\cdot{X}}$ gives exactly the $\star$-exponential (plane wave) for symmetric quantization with the $\vec{k}$-parametrization.

The composition of coordinates can be inferred from 
\bas
e^{i\vec{k}_1\cdot{X}}\star_\cs e^{i\vec{k}_2\cdot{X}} =\cs^{-1}(\cs(e^{i\vec{k}_1\cdot{X}})\cdot \cs(e^{i\vec{k}_2\cdot{X}}))=e^{i\cb(\vec{k}_1,\vec{k}_2)\cdot{X}}\,,
\eas
where $\cb(\vec{k}_1,\vec{k}_2)$ is the Baker-Campbell-Hausdorff formula (cf.\@ Appendix \ref{app:bch} for closed formula for $\SU(2)$), and therefore, the $\star_\cs$-product on $\star$-monomials can be computed according to (\ref{eq:monomialsstar}):
\bas
X_i\star_\cs X_j &= X_iX_j+i\eps_{ij}^{\ \ k}X_k\,,\\
X_i\star_\cs X_j \star_\cs X_k & =X_iX_jX_k + i (\eps_{ijm}X_k+\eps_{ikm}X_j+\eps_{jkm}X_i)X_m+\frac{2}{3}\delta_{jk}X_i-\frac{1}{3}\delta_{ik}X_j+\frac{2}{3}\delta_{ij}X_k\,,\\
\vdots\nonumber
\eas
This star-product is referred to as the Gutt (or \lq standard\rq) $\star$-product \cite{gutt}. As explained above, for plane waves we amend this product by a projection, which explicitly gives
\bas
e^{i\vec{k}_1\cdot{X}}\star_{\cs p} e^{i\vec{k}_2\cdot{X}} =e^{i\cb_p(\vec{k}_1,\vec{k}_2)\cdot{X}}\,,
\eas
where $\cb_p(\vec{k}_1,\vec{k}_2)$ is the value of the Baker-Campbell-Hausdoff formula projected onto the principal branch of the logarithm map. Under integration, using (\ref{eq:starint}) and (\ref{eq:kHaar}), the $\star_{\mc{S}p}$-product acquires the form
\bas
	\int_{\fg^*}\d^3X\ f(X) \star_{\mc{S}p} f'(X) = \int_{\fg^*}\d^3X\ f(X) \left(\frac{|\vec{\prt}|}{\sin|\vec{\prt}|}\right)^2 f'(X) \,.
\eas

Given the plane waves just computed, we may then write the explicit form for the non-commutative Fourier transform as
\ba
  \tilde{\psi}({X})= \int_{\Rl^3,|\vec{k}|\in[0,\pi[}\d^3k\, \left(\frac{\sin|\vec{k}|}{|\vec{k}|}\right)^2   \ e^{{i}\vec{k}\cdot {X}}\, \psi(\vec{k}) \,,
\ea
with the inverse, from (\ref{eq:Finvsigma}), being
\ba
  \psi(\vec{k}) = \left(\frac{|\vec{k}|}{\sin|\vec{k}|}\right)^2 \int_{\Rl^3} \frac{\d^3 {X}}{(2\pi)^3}\ e^{{-i}\vec{k}\cdot X}\, \tilde{\psi}({X}) \,.
\ea

\subsubsection{Duflo map}
The Duflo map, as defined in more detail in the Appendix \ref{app:uea} is given by
\bas
\mathcal{D}=\mathcal{S}\circ j^{\frac{1}{2}}(\del)\,,
\eas
where $j$ is the following function on $\fg$
\bas
j(X)=\det\left(\frac{\sinh \frac{1}{2}\text{ad}_X}{\frac{1}{2}\text{ad}_X}\right)\,.
\eas
For $X\in\su(2)$, $j$ computes to
\bas
j(X)=\left(\frac{\sinh |{X}|}{|{X}|}\right)^2\,.
\eas
The application of the Duflo quantization map to exponentials $e^{i\vec{k}\cdot {X}}$ gives
\bas
\mathcal{D}(e^{i\vec{k}\cdot{X}})=\frac{\sin |\vec{k}|}{|\vec{k}|}e^{i\vec{k}\cdot\hat{{X}}}\,, 
\eas
which can be inverted to give
\bas
\mathcal{D}^{-1}(e^{i\vec{k}\cdot\hat{{X}}})=\frac{|\vec{k}|}{\sin |{\vec{k}}|}\,e^{i\vec{k}\cdot{X}}\equiv  e_\star^{i\vec{k}\cdot{X}}\,,
\eas
that is, we have found the plane wave $E_g(X)$ under $\cd$ with the $\vec{k}$-parametrization. This result, as other aspects of our construction, extends and confirms from a different perspective, the derivation in \cite{Freidel:2005ec}.

Once again, we may now use (\ref{eq:monomialsstar}) to compute the $\star_\cd$-product on monomials:
\bas
X_i\star_\cd X_j &= X_iX_j+i\eps_{ij}^{\ \ k}X_k-\frac{1}{3}\delta_{ij}\,,\\
X_i\star_\cd X_j \star_\cd X_k & =X_iX_jX_k + i (\eps_{ijm}X_k+\eps_{ikm}X_j+\eps_{jkm}X_i)X_m+\frac{1}{3}\delta_{jk}X_i-\frac{2}{3}\delta_{ik}X_j+\frac{1}{3}\delta_{ij}X_k\,,\\
\vdots\nonumber
\eas
This star-product coincides with the star-product introduced by Kontsevich in \cite{kontsevich}. For the non-commutative plane wave we again have the corresponding projected star-product $\star_{\cd p}$, which satisfies
\bas
	\frac{|\vec{k}_1|}{\sin |{\vec{k}_1}|}\,e^{i\vec{k}_1\cdot{X}} \star_{\cd p} \frac{|\vec{k}_2|}{\sin |{\vec{k}_2}|}\,e^{i\vec{k}_2\cdot{X}} = \frac{|\cb_p(\vec{k}_1,\vec{k}_2)|}{\sin |{\cb_p(\vec{k}_1,\vec{k}_2)}|}\,e^{i\cb_p(\vec{k}_1,\vec{k}_2)\cdot{X}} \,.
\eas
Again, an expression for the $\star_{\cd p}$-product under integration can be obtained from (\ref{eq:starint}). However, for the Duflo map the factors $\omega$ and $\eta^2$ cancel out exactly, and we have $\sigma(\zeta)^{-1} \equiv \omega(\zeta)|\eta(\zeta)|^2 = 1$. Accordingly,
\bas
	\int_{\fg^*}\d^3X\ f(X) \star_{\cd p} f'(X) = \int_{\fg^*}\d^3X\ f(X) f'(X) \,,
\eas
i.e., the Duflo star-product coincides with the pointwise product (only) under integration. In particular, this implies that the Duflo $L_\star^2$ inner product coincides with the usual $L^2$ inner product, and therefore $L_\star^2(\fg^*) \subseteq L^2(\fg^*)$ (as an $L^2$ norm-complete vector space) for the Duflo map.

The explicit form of the non-commutative Fourier transform is thus
\ba
  \tilde{\psi}(X)= \int_{\Rl^3,|\vec{k}|\in[0,\pi[}\d^3k\, \left(\frac{\sin|\vec{k}|}{|\vec{k}|}\right)  \ e^{{i}\vec{k}\cdot {X}}\, \psi(\vec{k}) \,,
\ea
while the inverse is
\ba
  \psi(\vec{k}) = \int_{\Rl^3} \frac{\d^3 {X}}{(2\pi)^3}\ \left(\frac{|\vec{k}|}{\sin |\vec{k}|}\right) e^{{-i}\vec{k}\cdot {X}}\, \tilde{\psi}(X) \,.
\ea

\subsubsection{Freidel-Livine-Majid map}
The Freidel-Livine-Majid ordering map $\mathcal{Q}_{\text{FLM}}$ \cite{Freidel:2005ec}, which has found several applications in the quantum gravity literature (cited in the introduction), can be essentially seen as symmetrization map in conjunction with a change of parametrization for $\SU(2)$. In particular, for exponentials of the form $e^{i\vec{p}\cdot \vec{X}}$ it is defined as
\ba
\cq_{\text{FLM}}(e^{i\vec{p}\cdot {X}}):=e^{i\frac{\sin^{-1}|\vec{p}|}{|\vec{p}|}\vec{p}\cdot \hat{{X}}}\,,
\label{eq:quantmap}
\ea
which implies
\bas
\cq_{\text{FLM}}(e^{i\frac{\sin|\vec{k}|}{|\vec{k}|}\vec{k}\cdot {X}})=e^{i\vec{k}\cdot\hat{{X}}},
\eas
that is, with the $\vec{k}$-parametrization, the plane wave is given by $e_\star^{i\vec{k}\cdot{X}}=e^{i\frac{\sin|\vec{k}|}{|\vec{k}|}\vec{k}\cdot {X}}$. Accordingly, we have 
\bas
\mathcal{Q}_{\text{FLM}}^{-1}(e^{i\vec{k}\cdot\hat{{X}}})=e^{i\frac{\sin|\vec{k}|}{|\vec{k}|}\vec{k}\cdot{X}}\,.
\eas
Of course, the transformation $\frac{\sin|\vec{k}|}{|\vec{k}|}\vec{k}$ defines the $\vec{p}$-parametrization as of (\ref{eq:pparametrization}), and therefore we may simply write $e_\star^{i\vec{k}\cdot {X}}=e^{i\vec{p}(\vec{k})\cdot{X}}$. However, the coordinates $\vec{p}$ only cover the upper (or lower) hemisphere $\SU(2)/\Zl_2 \cong \SO(3)$, and the resulting non-commutative Fourier transform is applicable only for functions on $\SO(3)$.

Using the expression (\ref{bchsu2}) for the Baker-Campbell-Hausdorff formula for $\su(2)$ we have
\bas
e^{i\vec{p}_1\cdot{X}}\star_{\text{FLM}} e^{i\vec{p}_2\cdot{X}} &=\mathcal{Q}_{\text{FLM}}^{-1}(\mathcal{Q}_{\text{FLM}}(e^{i\vec{p}_1\cdot{X}})\cdot \mathcal{Q}_{\text{FLM}}(e^{i\vec{p}_2\cdot{X}})) \nn
&=\mathcal{Q}_{\text{FLM}}^{-1}\left(e^{i\frac{\sin^{-1}|\vec{p}_1|}{|\vec{p}_1|}\vec{p}_1\cdot{\hat{X}}}\cdot e^{i\frac{\sin^{-1}|\vec{p}_2|}{|\vec{p}_2|}\vec{p}_2\cdot{\hat{X}}}\right)\nn
&=\mathcal{Q}_{\text{FLM}}^{-1}\left(e^{i\cb\left(\frac{\sin^{-1}|\vec{p}_1|}{|\vec{p}_1|}\vec{p}_1,\frac{\sin^{-1}|\vec{p}_2|}{|\vec{p}_2|}\vec{p}_2\right)\cdot{\hat{X}}}\right) \nn
&=\mathcal{Q}_{\text{FLM}}^{-1}\left(e^{i \frac{\sin^{-1}|\vec{p}_1\oplus\vec{p}_2|}{|\vec{p}_1\oplus\vec{p}_2|}\vec{p}_1\oplus\vec{p}_2 \cdot {\hat{X}}}\right)\nn
&=e^{i(\vec{p}_1\oplus\vec{p}_2)\cdot{X}}\,,
\eas
where
\bas
\vec{p}_1\oplus\vec{p}_2=\sqrt{1-|\vec{p}_2|^2}\,\vec{p}_1+\sqrt{1-|\vec{p}_1|^2}\,\vec{p}_2-\vec{p}_1\times \vec{p}_2\,.
\eas
Now, since the $\vec{p}$-parametrization is applicable only for the upper hemisphere of $\SU(2)$, that is, $\SO(3)$, instead of restricting the parametrization of the non-commutative plane waves to the principal branch of the logarithm, we restrict to the upper hemisphere, and introduce the corresponding projection into the star-product of non-commutative plane waves as
\bas
e^{i\vec{p}_1\cdot{X}}\star_{\text{FLM}p} e^{i\vec{p}_2\cdot{X}} = e^{i(\vec{p}_1\oplus_p\vec{p}_2)\cdot{X}}\,,
\eas
where
\bas
\vec{p}_1\oplus_p\vec{p}_2=\eps(\vec{p}_1,\vec{p}_2)\left(\sqrt{1-|\vec{p}_2|^2}\,\vec{p}_1+\sqrt{1-|\vec{p}_1|^2}\,\vec{p}_2-\vec{p}_1\times \vec{p}_2\right)\,.
\eas
The factor $\eps(\vec{k}_1,\vec{k}_2)=\sgn(\sqrt{1-|\vec{p}_1|^2}\sqrt{1-|\vec{p}_2|^2}-\vec{p}_1\cdot\vec{p}_2)$, introduced by the projection, is 1 if both $\vec{p}_1,\vec{p}_2$ are close to zero or one of them is infinitesimal, and $-1$ when the addition of two upper hemisphere vectors ends up in the lower hemisphere, thus projecting the result to its antipode on the upper hemisphere.

The $\star_{\text{FLM}}$-monomials thus read
\bas
X_i\star_{\text{FLM}} X_j &= X_iX_j+i\eps_{ij}^{\ \ k}X_k\,,\\
X_i\star_{\text{FLM}} X_j \star_{\text{FLM}} X_k & =X_iX_jX_k + i (\eps_{ijm}X_k+\eps_{ikm}X_j+\eps_{jkm}X_i)X_m+\delta_{jk}X_i-\delta_{ik}X_j+\delta_{ij}X_k\,,\\
\vdots\nonumber
\eas
which coincide with $\star_\cs$ to second order, but no further.

As was already shown in \cite{Freidel:2005me,Livine:2008hz}, but rederivable from the general expression (\ref{eq:starint}) and (\ref{eq:pHaar}), for the Freidel-Livine-Majid star-product we have under integration
\bas
	\int_{\fg^*}\d^3X\ f(X) \star_{\text{FLM}p} f'(X) = \int_{\fg^*}\d^3X\ f(X)\, \sqrt{1+{\nabla}^2}\, f'(X) \,.
\eas

Now, given the plane waves just computed, we may write the explicit form of the non-commutative Fourier transform as
\ba
  \tilde{\psi}(X)=\int_{\Rl^3, |\vec{p}|^2<1}\frac{\d^3 {p}}{\sqrt{1-|\vec{p}|^2}}   \ e^{{i}\vec{p}\cdot {X}}\, \psi(\vec{p}) \,,
\ea
as well as the inverse
\ba
  \psi(\vec{p}) = \sqrt{1-|\vec{p}|^2} \int_{\Rl^3} \frac{\d^3 {X}}{(2\pi)^3}\ e^{{-i}\vec{p}\cdot {X}}\, \tilde{\psi}(X) \,.
\ea

\section{Conclusion}
\label{sec:conc}
We have studied the representations of the quantum algebra $\fA$ obtained by canonically quantizing the Poisson algebra $\cp_G$ associated to the cotangent bundle of a Lie group $G$ (with Lie algebra $\fg$). In addition to the usual representation of $\fA$ on the Hilbert space of square-integrable functions $L^2(G)$ on $G$ (with respect to the Haar measure $\d g$),  we have shown that a dual \emph{algebra representation} of $\fA$ in terms of a function space we denote as $L^2_\star(\fg^*)$ can be defined (and identified the conditions for its existence) by introducing a suitable $\star$-product, in the sense of deformation quantization \cite{defquant}, depending only on the chosen quantization map between $\cp_G$ and $\fA$. The non-commutative Fourier transform is then defined as the intertwining map between these two representations. We have seen that the explicit form of the non-commutative plane wave, and thus that of the transform, depends again only on the choice of a quantization map or, equivalently, a deformation quantization $\star$-product. In fact, in terms of the canonical coordinates (of the first kind) $k(g) = -i\ln(g)\in\fg$ on $G$ obtained through the logarithm map, the plane wave is shown to be given by the star-exponential
\bas
E_g(X)=e_\star^{ik(g)\cdot X} \,,
\eas
where $X\in \fg^*$, which can then be equivalently written as standard exponentials for some (a priori different) choice of coordinates on the group, also following from the choice of quantization map. 

Our results show that the possibility of a non-commutative algebra representation does not require the existence of the group representation, but only a choice of quantization map.  The algebra representation for the quantum system, in other words, can stand on its own feet. Of course, which representation is more convenient to use depends on the specific question being tackled, as different representations have different advantages.

The results also offer a new perspective on the non-commutative Fourier transform and some more insights into the various elements entering in its definition (e.g., the choice of coordinates), and lead to a prescription for how to define plane waves for generic quantization maps. This also clarifies the relation with the so-called quantum group Fourier transform of Majid, extending the work of Freidel \& Majid \cite{Freidel:2005ec}.

In general, for an arbitrary quantization map and corresponding $\star$-product, the necessary conditions for the existence of the algebra representation would not be satisfied. However, we have provided some explicit and non-trivial examples of the above construction, satisfying the necessary conditions, in the case $G=\SU(2)$, corresponding to three choices of quantization maps: the symmetric map, the Duflo map, and the so-called Freidel-Livine-Majid map (used in the quantum gravity literature).
For these examples, we have provided the corresponding $\star$-product, algebra representation and non-commutative plane waves explicitly.\\

Besides clarifying some aspects and the underlying logic of the construction of the algebra representation and of the non-commutative Fourier transform, we expect our results to have also interesting applications in the study of specific quantum systems arising from the quantization of the phase space we started from. In particular, we hope to have provided new tools to the development of quantum gravity models in the context of loop quantum gravity and group field theory. For example, a first application of our construction would be to study the flux representation of loop quantum gravity and the corresponding coherent states for the Duflo map, extending the work of \cite{carlosbiancadaniele, Oriti:2011ug}. In the same direction, the construction of a new 4d gravity model along the same lines as \cite{Baratin:2011hp} can now be performed for the algebra representation corresponding, again, to the Duflo map, and it would be very interesting to identify clearly the consequences for the resulting model of the nice mathematical properties of such a quantization map.

\acknowledgments{CG is supported by the Portuguese Science Foundation (\emph{Funda\c{c}\~ao para a Ci\^encia e a Tecnologia}) under research grant SFRH/BD/44329/2008, which he greatly acknowledges. DO acknowledges support from the A. von Humboldt Stiftung through a Sofja Kovalevskaja Prize. We are very grateful to Aristide Baratin for extensive and useful comments and discussions.}

\appendix
\section{Universal enveloping algebras}
\label{app:uea}
Let $V$ be an $n$-dimensional vector space over $\Kl$ ($\Rl$ or $\Cl$) with basis $\{e_i\}_{i=1,\ldots,n}$, and define the tensor algebra over $V$ as
\bas
T^\bullet(V):=\bigoplus_{k=0}^\infty V^{\otimes k}=\Kl\oplus V\oplus (V\otimes V)\oplus (V\otimes V \otimes V)\oplus \cdots\,,
\eas
where multiplication is simply defined by concatenation. A generic element $v\in T^\bullet(V)$ can be written as
\ba
v=v^0+v^ie_i+v^{ij}e_i\otimes e_j+v^{ijk}e_i\otimes e_j\otimes e_k+\cdots\,,
\label{eq:generic}
\ea
where $v^0,v^i,v^{ij},v^{ijk},\ldots\in \Kl$, $i,j,k,\ldots=1,\ldots,n$, with no conditions on the coefficients.

The \textit{symmetric algebra} of $V$, $\text{Sym}(V)$, is then defined as the quotient of the tensor algebra $T^\bullet(V)$ by the two-sided ideal generated by the set
\bas
\mathfrak{I}=\{v\otimes w-w\otimes v\: :\: v,w\in V\}\,.
\eas
In particular, notice that $\text{Sym}(V)$ is a commutative algebra, and it is actually isomorphic to the polynomial algebra $\Kl[e_1,\ldots,e_n]$. A generic element $v\in \text{Sym}(V)$ can be written the same way as in (\ref{eq:generic}) but this time the coefficients are completely symmetric, $v^{ij}=v^{(ij)}$, $v^{ijk}=v^{(ijk)}$, \ldots, identifying $\text{Sym}(V)$ with the algebra of symmetric tensors on $V$. 
As a polynomial we would have $p(x_1,\ldots,x_n)=v^0+v^ix_i+v^{ij}x_ix_j+v^{ijk}x_ix_jx_k+\cdots$, with indeterminates $x_1,\ldots,x_n\in \Kl$, $i,j,k,\ldots=1,\ldots,n$.

In case $V=\fg$, the Lie algebra of the Lie group $G$ with Lie bracket $[\cdot,\cdot]$, we can define the \textit{universal enveloping algebra} of $V$, $U(\fg)$, as the quotient of the tensor algebra $T^\bullet(\fg)$ by the two-sided ideal generated by the set
\bas
\mathfrak{I'}=\{v\otimes w-w\otimes v-[v,w] \: :\: v,w\in\fg\}\,,
\eas
that is, $U(\fg)=T^\bullet(\fg)/\mathfrak{I}'$. Naturally, $U(\fg)$ is a non-commutative algebra, and can be identified with the polynomial algebra $\Kl[x_1,\ldots,x_n]$ with indeterminates $x_1,\ldots,x_n$ satisfying the commutation relations $[x_i,x_j]=f_{ij}^{\ \ k}x_k$ inherited from the Lie algebra structure $[e_i,e_j]=f_{ij}^{\ \ k}e_k$. Note that for the case of an abelian Lie algebra $\fg$, for which the Lie bracket is identically zero, the universal enveloping algebra $U(\fg)$ coincides with the symmetric algebra $\text{Sym}(\fg)$. A generic element $v\in U(\fg)$ can still be written as (\ref{eq:generic}), however implementing the ideal $\mathfrak{I}'$ would involve the structure constants at length. Luckily, the following theorem gives a natural basis for $U(\fg)$.
\begin{theorem}[Poincar\'{e}-Birkhoff-Witt]
Let  $\{e_i\}_{i=1,\ldots,n}$ be an ordered basis for the Lie algebra $\fg$, the monomials
\bas
e_{1}^{m_1}\cdots e_{n}^{m_n}\,,
\eas
with $m_1,\ldots,m_n$ positive integers, form a basis for the universal enveloping algebra $U(\fg)$.
\label{thm:pbw}
\end{theorem}
Thus,
\bas
v=\sum_{m_1,\ldots, m_n\geq 0}v^{m_1\cdots m_n}e_{1}^{m_1} \cdots e_{n}^{m_n}\,,\q v^{m_1\cdots m_n}\in\Kl\,.
\eas
The crucial point about $U(\fg)$ is that this algebra can be naturally identified with the algebra of right-invariant differential operators (of all finite orders) on $G$, making it a natural ground for the algebra of the quantum theory. The left action of $G$ on itself gives a natural action on functions $(L_g f)(h)=f(gh)$, $g,h\in G$. In turn, for each $X\in\fg$ we have its action on functions as differential operators $(\cl_X f)(g)=\left.\frac{\d}{\d t}\right|_{t=0}f(e^{tX}g)$, thus identifying $\fg$ with the right-invariant vector fields on $G$, or rather the right-invariant differential operators of order one. Extending this inclusion to the full $U(\fg)$ gives the desired mapping. Furthermore, the center of $U(\fg)$, denoted $\cz(U(\fg))$, consists of the left- and right- invariant differential operators, of which the Casimir operators are a prime example.

\subsection{The Duflo map}
We may now define the \textit{symmetrization map} (or symmetric quantization):
\ba
\mathcal{S} :  \text{Sym}(\fg) &\longrightarrow U(\fg)\nonumber \\
 X_1\cdots X_k &\longmapsto \frac{1}{k!}\sum_{\sigma\in S_k} X_{\sigma_1}\cdots X_{\sigma_k}\,,
 \label{eq:symmetrizationmap}
\ea
where $S_k$ is the symmetric group of order $k$. On the other hand, the symmetrization map may be completely characterized by being the identity on $\fg$, linear, and satisfying the property $\cs(X^n)=\cs(X)^n$ for all $X\in\fg$, and $n\geq 0$. The idea of $\cs$ is to map as surjectively as possible a commutative algebra to a non-commutative algebra, and it is obviously \textit{not} an algebra isomorphism unless $\fg$ is abelian (though it can be proved to be a \textit{linear} isomorphism).

Invariant polynomials, i.e., elements of $\text{Sym}(\fg)$ invariant under the (adjoint) action of $G$, denoted $\text{Sym}(\fg)^\fg$, are particularly important since they map to Casimirs under any quantization scheme. In fact, there exists an algebra isomorphism between the subalgebras $\text{Sym}(\fg)^\fg$ and $U(\fg)^\fg$, the latter corresponding to the $G$-invariant differential operators on $U(\fg)$ (which is an alternative definition for the center of $U(\fg)$, that is, $U(\fg)^\fg=\cz(U(\fg))$). The map giving such an isomorphism is called the \textit{Duflo map} (or Duflo quantization) and is given explicitly by
\ba
\mathcal{D}=\mathcal{S}\circ j^{\frac{1}{2}}(\del)\,,
\label{eq:duflomap}
\ea
where $j$ is the following function on $\fg$\footnote{It is curious to note that the function $j$ appears also in other contexts. (1) Changing measure from the Lie group $G$ to the Lie algebra $\fg$: $\d g=j(X)\d X$, where $g=\exp X$. (2) Kirillov's character formula: 
\ba
\text{Tr}\,\pi_\l(\exp X)=\frac{1}{j^{1/2}(X)} \protect\int_{\co_{\l+\rho}}\d\mu_{\co_{\l+\rho}}(\xi)\,e^{i\langle \xi,X\rangle}\,,
\ea
where $\pi_\lambda$ is the unirrep for $\lambda\in\protect\widehat{G}$, $\rho$ is the half sum of the positive roots, $\co_{\l+\rho}$ the orbit passing through the point $\xi=i^{-1}(\l+\rho)\in \fg^*$, and $\d\mu_{\co_{\l+\rho}}(\xi)$ is a $G$-invariant measure on $\co_{\l+\rho}$.}
\ba
j(X)=\det\left(\frac{\sinh \frac{1}{2}\text{ad}_X}{\frac{1}{2}\text{ad}_X}\right)\,.
\label{eq:duflofactor}
\ea
Physically, the Duflo map tells that the centers of the ``classical" and ``quantum" level are the same. For semisimple Lie algebras $\fg$ the Duflo map coincides with the Harish-Chandra isomorphism. 

Finally, we note that the modified Duflo factor $\tilde{j}(X)=\det\left(\frac{1-e^{-\text{ad}_X}}{\text{ad}_X}\right)$ also gives the same algebra isomorphism. The one parameter group of automorphisms of $\text{Sym}(\fg)$ associated with the series
\bas
X\longmapsto \exp(\text{const}\cdot\text{Tr}(\text{ad}_X))
\eas
preserves the structure of the Poisson algebra on $\fg^*$, and indeed $\tilde{j}(X)=\det\left(e^{-\text{ad}_X/2}\right)j(X)=e^{-\text{Tr}(\text{ad}_X)/2}j(X)=j(X)$. It would, thus, be interesting to investigate further the unicity of the Duflo map, at least, in the restricted case of semisimple Lie algebras.

\section{On a property of the $\star_p$-product under integration}
\label{app:starint}
In this appendix we prove the identity (\ref{eq:starint}) stated without proof in the main text. Let us first note that
\bas
	\delta(g^{-1}h) = \omega(\zeta(h))^{-1} \delta^d(\zeta(g) - \zeta(h))\,,
\eas
where the first delta function is the one with respect to the Haar measure on the Lie group $G$, and the second one is the delta function with respect to the Lebesgue measure on the Euclidean space $\fg\cong\Rl^d$ of the coordinates $\zeta$. The proportionality is given by the inverse of the measure factor $\omega(\zeta)$, which gives the Haar measure in terms of the Lebesgue measure as $\d g = \omega(\zeta(g))\d \zeta(g)$. This can be checked by noting that
\bas
	f(g) &= \int_{G} \d h\ f(h)\, \delta(g^{-1}h) \nonumber\\
	&= \int_{G} \omega(\zeta(h)) \d \zeta(h)\ \hat{f}(\zeta(h))\, \omega(\zeta(h))^{-1}\, \delta^d(\zeta(g) - \zeta(h)) \quad \Big( = \hat{f}(\zeta(g)) \Big) \,,
\eas
where $f =: \hat{f} \circ \zeta$. Accordingly, we have
\bas
	\int_{\fg^*} \d^d X\ \overline{E_{g}(X)} \star_p E_h(X) &= (2\pi)^d \delta(g^{-1}h) \nn
	&= \omega(\zeta(h))^{-1}\, (2\pi)^d \delta^d(\zeta(g) - \zeta(h)) \nn
	&= \omega(\zeta(h))^{-1} \int_{\fg^*}\d^d X\ e^{-i\zeta(g) \cdot X} e^{i\zeta(h) \cdot X} \nn
	&= \omega(\zeta(h))^{-1} |\eta(\zeta(h))|^{-2} \int_{\fg^*}\d^d X\ \eta(-\zeta(g)) e^{-i\zeta(g) \cdot X} \eta(\zeta(h)) e^{i\zeta(h) \cdot X} \,,
\eas
where $\eta(\zeta(g)) := E(g,0)$, and in the last equality we used $\eta(-\zeta) = \overline{\eta(\zeta)}$ and the fact that the expression is non-zero only for $\zeta(g)=\zeta(h)$. But here the integrand is exactly a product of two non-commutative plane waves, and the prefactor we may write as a differential operator acting on one of the plane waves as
\bas
	\omega(\zeta(h))^{-1} |\eta(\zeta(h))|^{-2} E_h(X) = \omega(-i\vec{\prt})^{-1} |\eta(-i\vec{\prt})|^{-2} E_h(X) \,,
\eas
or, alternatively, as
\bas
	\omega(\zeta(g))^{-1} |\eta(\zeta(g))|^{-2} \overline{E_g(X)} = \omega(i\vec{\prt})^{-1} |\eta(i\vec{\prt})|^{-2} \overline{E_g(X)} \,.
\eas
We therefore have
\bas
	\int_{\fg^*} \d^d X\ \overline{E_{g}(X)} \star_p E_h(X) &= \int_{\fg^*} \d^d X\ \left( \big((\omega|\eta|^2)(i\vec{\prt}) \big)^{-1}\overline{E_g(X)} \right) E_h(X) \nn
	&= \int_{\fg^*} \d^d X\ \overline{E_g(X)} \left( \big((\omega|\eta|^2)(-i\vec{\prt}) \big)^{-1} E_h(X)\right) \,.
\eas
Linearity gives the sought for property (\ref{eq:starint}).

\section{Closed Baker-Campbell-Hausdorff formula for $\SU(2)$}
\label{app:bch}
Using the properties of the Pauli matrices $\sigma_i$ ($i=1,2,3$) we have the following expansion
\bas
g_j =e^{i \vec{k}_j\cdot \vec{\sigma}}=\cos|\vec{k}_j|\1_2+i\frac{\sin|\vec{k}_j|}{|\vec{k}_j|}\vec{k}_j\cdot\vec{\sigma}\,,\q (j=1,2)
\eas
which on multiplying two elements explicitly gives
\ba
g_1g_2 &=\left(\cos|\vec{k}_1|\cos|\vec{k}_2|-\frac{\sin|\vec{k}_1|\sin|\vec{k}_2|}{|\vec{k}_1||\vec{k}_2|}\vec{k}_1\cdot\vec{k}_2\right)\1_2 \nonumber\\
&+i\left(\frac{\cos|\vec{k}_2|\sin|\vec{k}_1|}{|\vec{k}_1|}\vec{k}_1
+\frac{\cos|\vec{k}_1|\sin|\vec{k}_2|}{|\vec{k}_2|}\vec{k}_2
-\frac{\sin|\vec{k}_1|\sin|\vec{k}_2|}{|\vec{k}_1||\vec{k}_2|}\vec{k}_1\times\vec{k}_2\right)\cdot\vec{\sigma}\,.
\label{eq:bch1}
\ea
The Baker-Campbell-Hausdorff formula is defined by the product of two exponentials
\bas
g_1g_2 =e^{i \vec{k}_1\cdot \vec{\sigma}}e^{i \vec{k}_2\cdot \vec{\sigma}}=e^{i \cb(\vec{k}_1,\vec{k}_2)\cdot \vec{\sigma}}
\eas
with a series expansion given by
\bas
\cb(\vec{k}_1,\vec{k}_2)=\vec{k_1}+\vec{k}_2-\vec{k}_1\times\vec{k}_2+\frac{1}{3}\vec{k}_1\times(\vec{k}_1\times\vec{k}_2)+\cdots\,.
\eas
Again by the properties of the Pauli matrices we have an analogous formula
\ba
g_1g_2 =\cos|\cb(\vec{k_1},\vec{k}_2)|\1_2+i\frac{\sin|\cb(\vec{k_1},\vec{k}_2)|}{|\cb(\vec{k_1},\vec{k}_2)|}\cb(\vec{k_1},\vec{k}_2)\cdot\vec{\sigma} \,.
\label{eq:bch2}
\ea
Identifying the appropriate terms in (\ref{eq:bch1}) and (\ref{eq:bch2}) we obtain the desired expression
\bas
\cb(\vec{k_1},\vec{k}_2)
=&\frac{\cos^{-1}\left(\cos|\vec{k}_1|\cos|\vec{k}_2|-\frac{\sin|\vec{k}_1|\sin|\vec{k}_2|}{|\vec{k}_1||\vec{k}_2|}\vec{k}_1\cdot\vec{k}_2\right)}{\sin\cos^{-1}\left(\cos|\vec{k}_1|\cos|\vec{k}_2|-\frac{\sin|\vec{k}_1|\sin|\vec{k}_2|}{|\vec{k}_1||\vec{k}_2|}\vec{k}_1\cdot\vec{k}_2\right)}\nn
&\times\left(\frac{\cos|\vec{k}_2|\sin|\vec{k}_1|}{|\vec{k}_1|}\vec{k}_1 
+\frac{\cos|\vec{k}_1|\sin|\vec{k}_2|}{|\vec{k}_2|}\vec{k}_2
-\frac{\sin|\vec{k}_1|\sin|\vec{k}_2|}{|\vec{k}_1||\vec{k}_2|}\vec{k}_1\times\vec{k}_2\right)\,.
\eas
Writing $\vec{k}_j$ as $\frac{\sin^{-1}|\vec{p}_j|}{|\vec{p}_j|}\vec{p}_j$ the formula can be expressed as a deformed addition of $\vec{p}_j$'s
\ba
\cb\left(\frac{\sin^{-1}|\vec{p}_1|}{|\vec{p}_1|}\vec{p}_1,\frac{\sin^{-1}|\vec{p}_2|}{|\vec{p}_2|}\vec{k}_2\right)
&=\frac{\cos^{-1}\sqrt{1-|\vec{p}_1\oplus\vec{p}_2|^2}}{\sin\cos^{-1}\sqrt{1-|\vec{p}_1\oplus\vec{p}_2|^2}}\,\vec{p}_1\oplus\vec{p}_2 \nonumber\\
&=\frac{\sin^{-1}|\vec{p}_1\oplus\vec{p}_2|}{|\vec{p}_1\oplus\vec{p}_2|}\,\vec{p}_1\oplus\vec{p}_2\,,
\label{bchsu2}
\ea
where $\vec{p}_1\oplus \vec{p}_2$ is given by
\bas
\vec{p}_1\oplus\vec{p}_2=\sqrt{1-|\vec{p}_2|^2}\,\vec{p}_1+\sqrt{1-|\vec{p}_1|^2}\,\vec{p}_2-\vec{p}_1\times \vec{p}_2\,,
\eas
and we have used
\bas
\sin\cos^{-1}x=\sqrt{1-x^2}=\cos\sin^{-1}x\,.
\eas

\bibliographystyle{JHEP}
\bibliography{thebibliography}
\end{document}

%% file: preamblejhep.tex
\usepackage{bbm}
\usepackage{graphicx}
\usepackage{latexsym}
\usepackage{color}

\usepackage{amsmath,amsthm,amssymb,amsfonts,amscd}
\newtheorem{theorem}{Theorem}[section]

\numberwithin{equation}{section}

\def\ba#1\ea{\begin{align}#1\end{align}}
\def\bas#1\eas{\begin{align*}#1\end{align*}}
\newcommand{\mt}{\mapsto}

\newcommand\q{\quad}
\newcommand{\del}{\partial}
\newcommand{\1}{\mathbbm{1}}
\renewcommand{\d}{{\rm d}}

\newcommand{\Nl}{\mathbb N}
\newcommand{\Zl}{\mathbb Z}

\newcommand{\Rl}{\mathbb R}
\newcommand{\Cl}{\mathbb C}

\newcommand{\Kl}{\mathbb K}

\newcommand{\ca}{\mathcal A}
\newcommand{\cb}{\mathcal B}
\newcommand{\cc}{\mathcal C}
\newcommand{\cd}{\mathcal D}
\newcommand{\ce}{\mathcal E}
\newcommand{\cf}{\mathcal F}

\newcommand{\ch}{\mathcal H}
\newcommand{\ci}{\mathcal I}

\newcommand{\cl}{\mathcal L}

\newcommand{\cn}{\mathcal N}
\newcommand{\co}{\mathcal O}
\newcommand{\cp}{\mathcal P}
\newcommand{\cq}{\mathcal Q}

\newcommand{\cs}{\mathcal S}

\newcommand{\cz}{\mathcal Z}

     \newcommand{\fA}{\mathfrak{A}}

     \newcommand{\fF}{\mathfrak{F}}
\newcommand{\fg}{\mathfrak{g}}     
     \newcommand{\fH}{\mathfrak{H}}

\newcommand{\fu}{\mathfrak{u}}

\newcommand{\eps}{\epsilon}

\renewcommand{\l}{\lambda}

\newcommand{\SO}{{\rm SO}}

\newcommand{\su}{\mathfrak{su}}
\newcommand{\SU}{{\rm SU}}
\newcommand{\SL}{{\rm SL}}
\newcommand{\U}{{\rm U}}

\newcommand*{\prt}{\partial}
\newcommand*{\be}{\begin{equation}}
\newcommand*{\ee}{\end{equation}}
\newcommand*{\bea}{\begin{eqnarray}}
\newcommand*{\eea}{\end{eqnarray}}
\newcommand*{\bd}{\begin{displaymath}}
\newcommand*{\ed}{\end{displaymath}}
\newcommand*{\nn}{\nonumber\\}

\newcommand*{\mc}{\mathcal}

\newcommand*{\sgn}{\textrm{sgn}}

\newcommand*{\norm}[1]{{\left|\!\left|{#1}\right|\!\right|}} 

%% file: fourier_v2.bbl
\providecommand{\href}[2]{#2}\begingroup\raggedright\begin{thebibliography}{10}

\bibitem{TTbook}
T.~Thiemann, {\em Modern Canonical Quantum General Relativity}.
\newblock Cambridge University Press, 2007.

\bibitem{CRbook}
C.~Rovelli, {\em Quantum Gravity}.
\newblock Cambridge University Press, 2006.

\bibitem{Rovelli:2011eq}
C.~Rovelli, {\it {Zakopane lectures on loop gravity}},  {\em PoS} {\bf
  QGQGS2011} (2011) 003, [\href{http://xxx.lanl.gov/abs/1102.3660}{{\tt
  arXiv:1102.3660}}].

\bibitem{Perez:2012wv}
A.~Perez, {\it The spin-foam approach to quantum gravity},  {\em Living Rev.
  Relativity} {\bf 16} (2013), no.~3
  [\href{http://xxx.lanl.gov/abs/1205.2019}{{\tt arXiv:1205.2019}}].

\bibitem{Oriti:2011jm}
D.~Oriti, {\it {The microscopic dynamics of quantum space as a group field
  theory}},  {\em in ``Foundations of Space and Time", G. Ellis, J.Marugan,
  A.Weltman (eds.), Cambridge University Press} (2012) 257--320,
  [\href{http://xxx.lanl.gov/abs/1110.5606}{{\tt arXiv:1110.5606}}].
  AEI-2010-043.

\bibitem{Oriti:2006se}
D.~Oriti, {\it {The group field theory approach to quantum gravity}},  {\em in
  ``Approaches to Quantum Gravity", D.Oriti (editor), Cambridge University
  Press, Cambridge} (2009) [\href{http://xxx.lanl.gov/abs/gr-qc/0607032}{{\tt
  gr-qc/0607032}}].

\bibitem{Baratin:2011aa}
A.~Baratin and D.~Oriti, {\it {Ten questions on group field theory (and their
  tentative answers)}},  {\em J. Phys. Conf. Ser.} {\bf 360} (2012) 012002,
  [\href{http://xxx.lanl.gov/abs/1112.3270}{{\tt arXiv:1112.3270}}].

\bibitem{Freidel:2005bb}
L.~Freidel and E.~R. Livine, {\it {Ponzano-Regge model revisited III: Feynman
  diagrams and effective field theory}},  {\em Class. Quant. Grav.} {\bf 23}
  (2006) 2021--2062, [\href{http://xxx.lanl.gov/abs/hep-th/0502106}{{\tt
  hep-th/0502106}}].

\bibitem{Freidel:2005ec}
L.~Freidel and S.~Majid, {\it {Noncommutative harmonic analysis, sampling
  theory and the Duflo map in 2+1 quantum gravity}},  {\em Class. Quant. Grav.}
  {\bf 25} (2008) 045006, [\href{http://xxx.lanl.gov/abs/hep-th/0601004}{{\tt
  hep-th/0601004}}].

\bibitem{JoungMouradNoui}
E.~Joung, J.~Mourad, and K.~Noui, {\it Three dimensional quantum geometry and
  deformed {P}oincare symmetry},  {\em J. Math. Phys.} {\bf 50} (2009) 052503,
  [\href{http://xxx.lanl.gov/abs/0806.4121}{{\tt arXiv:0806.4121}}].

\bibitem{Baratin:2010wi}
A.~Baratin and D.~Oriti, {\it {Group field theory with non-commutative metric
  variables}},  {\em Phys. Rev. Lett.} {\bf 105} (2010) 221302,
  [\href{http://xxx.lanl.gov/abs/1002.4723}{{\tt arXiv:1002.4723}}].

\bibitem{Baratin:2011tx}
A.~Baratin and D.~Oriti, {\it {Quantum simplicial geometry in the group field
  theory formalism: Reconsidering the Barrett-Crane model}},  {\em New J.
  Phys.} {\bf 13} (2011) 125011, [\href{http://xxx.lanl.gov/abs/1108.1178}{{\tt
  arXiv:1108.1178}}].

\bibitem{Baratin:2011hp}
A.~Baratin and D.~Oriti, {\it {Group field theory and simplicial gravity path
  integrals: A model for Holst-Plebanski gravity}},  {\em Phys. Rev.} {\bf D85}
  (2012) 044003, [\href{http://xxx.lanl.gov/abs/1111.5842}{{\tt
  arXiv:1111.5842}}].

\bibitem{Baratin:2011tg}
A.~Baratin, F.~Girelli, and D.~Oriti, {\it {Diffeomorphisms in group field
  theories}},  {\em Phys. Rev.} {\bf D83} (2011) 104051,
  [\href{http://xxx.lanl.gov/abs/1101.0590}{{\tt arXiv:1101.0590}}].

\bibitem{Oriti:2011ug}
D.~Oriti, R.~Pereira, and L.~Sindoni, {\it {Coherent states in quantum gravity:
  a construction based on the flux representation of LQG}},  {\em J. Phys.}
  {\bf A45} (2012) 244004, [\href{http://xxx.lanl.gov/abs/1110.5885}{{\tt
  arXiv:1110.5885}}].

\bibitem{DanieleMatti}
D.~Oriti and M.~Raasakka, {\it Quantum mechanics on {SO(3)} via non-commutative
  dual variables},  {\em Phys. Rev.} {\bf D84} (2011) 025003,
  [\href{http://xxx.lanl.gov/abs/1103.2098}{{\tt arXiv:1103.2098}}].

\bibitem{Dupuis:2011fx}
M.~Dupuis, F.~Girelli, and E.~R. Livine, {\it {Spinors and Voros star-product
  for group field theory: First contact}},  {\em Phys. Rev.} {\bf D86} (2012)
  105034, [\href{http://xxx.lanl.gov/abs/1107.5693}{{\tt arXiv:1107.5693}}].

\bibitem{Alekseev:2000hf}
A.~Alekseev, A.~Polychronakos, and M.~Smedback, {\it {On area and entropy of a
  black hole}},  {\em Phys. Lett.} {\bf B574} (2003) 296--300,
  [\href{http://xxx.lanl.gov/abs/hep-th/0004036}{{\tt hep-th/0004036}}].

\bibitem{Sahlmann:2011rv}
H.~Sahlmann and T.~Thiemann, {\it {Chern-Simons expectation values and quantum
  horizons from LQG and the Duflo map}},  {\em Phys. Rev. Lett.} {\bf 108}
  (2012) 111303, [\href{http://xxx.lanl.gov/abs/1109.5793}{{\tt
  arXiv:1109.5793}}].

\bibitem{Sahlmann:2011uh}
H.~Sahlmann and T.~Thiemann, {\it {Chern-Simons theory, Stokes' theorem, and
  the Duflo map}},  {\em J. Geom. Phys.} {\bf 61} (2011) 1104--1121,
  [\href{http://xxx.lanl.gov/abs/1101.1690}{{\tt arXiv:1101.1690}}].

\bibitem{Noui:2011im}
K.~Noui, A.~Perez, and D.~Pranzetti, {\it {Canonical quantization of
  non-commutative holonomies in 2+1 loop quantum gravity}},  {\em JHEP} {\bf
  1110} (2011) 036, [\href{http://xxx.lanl.gov/abs/1105.0439}{{\tt
  arXiv:1105.0439}}].

\bibitem{Schroers}
B.~J. {Schroers}, {\it {Combinatorial quantisation of Euclidean gravity in
  three dimensions}},  \href{http://xxx.lanl.gov/abs/math/0006228}{{\tt
  math/0006228}}.

\bibitem{Majid:2008iz}
S.~Majid and B.~Schroers, {\it {q-Deformation and semidualisation in 3d quantum
  gravity}},  {\em J. Phys.} {\bf A42} (2009) 425402,
  [\href{http://xxx.lanl.gov/abs/0806.2587}{{\tt arXiv:0806.2587}}].

\bibitem{Rosa:2012pr}
L.~Rosa and P.~Vitale, {\it {On the $\star$-product quantization and the Duflo
  map in three dimensions}},  {\em Mod. Phys. Lett.} {\bf A27} (2012) 1250207,
  [\href{http://xxx.lanl.gov/abs/1209.2941}{{\tt arXiv:1209.2941}}].

\bibitem{GraciaBondia:2001ct}
J.~M. Gracia-Bondia, F.~Lizzi, G.~Marmo, and P.~Vitale, {\it {Infinitely many
  star products to play with}},  {\em JHEP} {\bf 0204} (2002) 026,
  [\href{http://xxx.lanl.gov/abs/hep-th/0112092}{{\tt hep-th/0112092}}].

\bibitem{kirillov}
A.~A. Kirillov, {\em Lectures on the Orbit Method}.
\newblock American Mathematical Society, 2004.

\bibitem{wildberger}
N.~J. Wildberger, {\it On the {F}ourier transform of a compact semisimple {L}ie
  group},  {\em Journal of the Australian Mathematical Society (Series A)
  (Series A)} {\bf 56} (1994), no.~01 64--116.

\bibitem{helgason}
S.~Helgason, {\em Geometric analysis on symmetric spaces}, vol.~39 of {\em
  Mathematical Surveys and Monographs}.
\newblock American Mathematical Society, Providence, RI, second~ed., 2008.

\bibitem{defquant}
F.~Bayen, M.~Flato, C.~Fronsdal, A.~Lichnerowicz, and D.~Sternheimer, {\it
  Quantum mechanics as a deformation of classical mechanics},  {\em Lett. Math.
  Phys.} {\bf 1} (1977) 521--530. 10.1007/BF00399745.

\bibitem{rigged}
A.~B{\"o}hm, {\em The Rigged {H}ilbert Space and Quantum Mechanics}.
\newblock Lecture Notes in Physics. Springer Verlag, 1978.

\bibitem{nst}
I.~M. Gelfand and N.~Y. Vilenkin, {\em Generalized Functions - Vol 4:
  Applications of Harmonic Analysis}.
\newblock Academic Press, 1964.

\bibitem{obstruction}
M.~J. {Gotay}, H.~B. {Grundling}, and G.~M. {Tuynman}, {\it {Obstruction
  results in quantization theory}},  {\em J. NonLinear Sci.} {\bf 6} (Sept.,
  1996) 469--498, [\href{http://xxx.lanl.gov/abs/math-ph/9809011}{{\tt
  math-ph/9809011}}].

\bibitem{vN1}
J.~Neumann, {\it {Die Eindeutigkeit der Schr\"odingerschen Operatoren}},  {\em
  Mathematische Annalen} {\bf 104} (1931) 570--578.

\bibitem{vN2}
J.~Neumann, {\it {Ueber einen Satz von Herrn M. H. Stone}},  {\em Ann. Math.}
  {\bf 33} (1932) 567--573.

\bibitem{expstatus}
D.~Djokovic and K.~H. Hofmann, {\it The surjectivity question for the
  exponential function of real {L}ie groups: A status report},  {\em Journal of
  Lie Theory} {\bf 7} (1997) 171 -- 199.

\bibitem{explatest}
M.~Wuestner, {\it Lie groups with surjective exponential function},  {\em
  Shaker-Verlag, Aachen} (2001).

\bibitem{Vilasi:2001bm}
G.~Vilasi, {\em {Hamiltonian Dynamics}}.
\newblock World Scientific, 2001.

\bibitem{carlosbiancadaniele}
B.~Dittrich, C.~Guedes, and D.~Oriti, {\it {On the space of generalized fluxes
  for loop quantum gravity}},  {\em Class. Quant. Grav.} {\bf 30} (2013)
  055008, [\href{http://xxx.lanl.gov/abs/1205.6166}{{\tt arXiv:1205.6166}}].

\bibitem{gutt}
S.~{Gutt}, {\it {An explicit $^{*}$-product on the cotangent bundle of a Lie
  group}},  {\em Lett. Math. Phys.} {\bf 7} (May, 1983) 249--258.

\bibitem{kontsevich}
M.~Kontsevich, {\it Deformation quantization of {P}oisson manifolds},  {\em
  Lett. Math. Phys.} {\bf 66} (2003) 157--216.

\bibitem{Freidel:2005me}
L.~Freidel and E.~R. Livine, {\it {Effective 3-D quantum gravity and
  non-commutative quantum field theory}},  {\em Phys. Rev. Lett.} {\bf 96}
  (2006) 221301, [\href{http://xxx.lanl.gov/abs/hep-th/0512113}{{\tt
  hep-th/0512113}}].

\bibitem{Livine:2008hz}
E.~R. Livine, {\it {Matrix models as non-commutative field theories on R**3}},
  {\em Class. Quant. Grav.} {\bf 26} (2009) 195014,
  [\href{http://xxx.lanl.gov/abs/0811.1462}{{\tt arXiv:0811.1462}}].

\end{thebibliography}\endgroup
